\documentclass[aps,prd,twocolumn,showpacs]{revtex4}
\usepackage{graphicx}

\begin{document}

\title{Efficient Cosmological Parameter Estimation from  
Microwave Background Anisotropies}
\author{Arthur Kosowsky}
 \altaffiliation[Also at ]{
School of Natural Sciences, Institute for Advanced Study,
Einstein Drive, Princeton, New Jersey~~08540
}
 \email{kosowsky@physics.rutgers.edu}
\author{Milos Milosavljevic}
 \email{milos@physics.rutgers.edu}
\author{Raul Jimenez}
 \email{raulj@physics.rutgers.edu}
\affiliation{
Department of Physics and Astronomy, Rutgers University,
136 Frelinghuysen Road, Piscataway, New Jersey~~08854-8019
}

\date{May, 2002}

\begin{abstract}
We revisit the issue of cosmological parameter estimation in light of
current and upcoming high-precision measurements of the cosmic
microwave background power spectrum. Physical quantities which
determine the power spectrum are reviewed, and their connection to
familiar cosmological parameters is explicated. We present a set of
physical parameters, analytic functions of the usual cosmological
parameters, upon which the microwave background power spectrum depends
linearly (or with some other simple dependence) over a wide range of
parameter values. With such a set of parameters, microwave background
power spectra can be estimated with high accuracy and negligible
computational effort, vastly increasing the efficiency of cosmological
parameter error determination. The techniques presented here allow
calculation of microwave background power spectra $10^5$ times
faster than comparably accurate direct codes (after precomputing
a handful of power spectra). We discuss various issues of parameter
estimation, including parameter degeneracies, numerical precision,
mapping between physical and cosmological
parameters, and systematic errors, and illustrate these
considerations with an idealized model of the MAP experiment.
\end{abstract}


\maketitle

\section{Introduction}

By January 2003, microwave maps of the full sky at 0.2$^\circ$
resolution will be available to the world, the harvest of the
remarkable MAP satellite currently taking data \cite{MAP}.  The
angular power spectrum of the temperature fluctuations in these maps
will be determined to high precision on angular scales from the
resolution limit up to a dipole variation; this corresponds to about
800 statistically independent power spectrum measurements. Recent
measurements have given a taste of the data to come, although covering
much smaller patches of the sky and with potentially more serious
systematic errors \cite{dib00,han00,hal02}.

The main science driving these spectacular technical feats is the
determination of basic cosmological parameters describing our
Universe, and resulting insights into fundamental physics; see
\cite{kk99,hd02} for recent reviews and \cite{kos01} for a pedagogical
introduction.  The expected series of acoustic peaks in the microwave
background power spectrum encode enough information to make possible
the determination of numerous cosmological parameters {\it
simultaneously} \cite{jun96}. These parameters include the
long-sought Hubble parameter $H_0$, the large-scale geometry of the
Universe $\Omega$, the mean density of baryons in the Universe
$\Omega_b$, and the value of the mysterious but now widely accepted
cosmological constant $\Lambda$, along with parameters describing the
tiny primordial perturbations which grew into present structures, and
the redshift at which the Universe reionized due to the formation of
the first stars or other compact objects.  While most of these
parameters and other information will be determined to high precision,
one near-exact degeneracy and other approximate degeneracies exist
between these parameters \cite{bon94,bon99}.

The exciting prospects of providing definitive answers to some of
cosmology's oldest questions raise a potentially difficult technical
issue, namely finding constraints on a large-dimensional parameter
space. Given a set of data, what range of points in parameter space
give models with an acceptably good fit to the data? The answer
requires evaluating a likelihood function at many points in parameter
space; in particular, finding confidence regions in multidimensional
parameter space requires looking around in the space. This is a 
straightforward process, in principle. But a parameter space
with as many as ten dimensions requires evaluating a lot of models: a
grid with a crude 10 values per parameter contains ten billion
models. A direct calculation of this many models is prohibitive, even
with very fast computers. While some of the parameters are independent
of the others (i.e.  tensor modes), reducing the effective
dimensionality, many of the parameters will be constrained quite
tightly, requiring a finer sampling in that direction of parameter
space. An additional problem with brute-force grid-based methods is a
lack of flexibility: if additional parameters are required to
correctly describe the Universe, vast amounts of recalculation must be
done. Grid-based techniques have been used for analysis of
cosmological parameters from microwave background (e.g., \cite{gor94,teg00})
but clearly this method is not fast or flexible enough to deal with
upcoming data sets adequately.

A more sophisticated approach is to perform a search
of parameter space in the region of interest.  Relevant techniques are
well known and have been applied to the microwave background
\cite{dod00,mel00,chr01}. 
Reliable estimates of the error region in cosmological
parameter space can be obtained with random sets of around $10^5$
models, reducing the computational burden by a factor of 1000 or more
compared to the cruder grid search methods. On fast parallel computers
it is possible to compute the power spectrum for a given model in a
computation time on the order of a second \cite{spergel}, making Monte
Carlo error determinations feasible. For improved efficiency, useful
implementations do not recalculate the entire spectrum at each point
in parameter space, but rather use approximate power spectra based on
smaller numbers of calculated models \cite{kno01}.

Here we expand and refine this idea by presenting a set of
parameters, functions of the usual cosmological parameters, which
reflect the underlying physical effects determining the microwave
background power spectrum and thus result in particularly simple and
intuitive parameter dependences. Previous crude implementations
of this idea applied to low-resolution measurements of the
power spectrum \cite{dod95}; the current and upcoming power
spectrum measurements requires a far more refined implementation. 
Our set of parameters can be used
to construct computationally trivial but highly accurate approximate
power spectra, and large Monte Carlo computations can then be
performed with great efficiency.  Additional advantages of a
physically-based parameter set are that the degeneracy structure of
the parameter space can be seen much more clearly, and the Monte Carlo
itself takes significantly fewer models to converge.

Earlier Fisher-matrix approximations of parameter errors are a rough
implementation of this general idea: if the the power spectrum varies
exactly linearly with each parameter in some parameter set, and the
measurement errors are gaussian random distributed and uncorrelated,
then the likelihood function as a function of the parameters can be
computed {\it exactly} from the partial derivatives of the model with
respect to the parameters \cite{jun96,zal97a,bon97}.  Even if the
linear parameter dependence does not hold throughout the parameter
space, it will almost always be valid in some small enough region,
being the lowest-order term in a Taylor expansion. If this region is
at least as large as the resulting error region, then the Fisher
matrix provides a self-consistent approximation for error
determination. The difficulty is in finding a suitable parameter
set. Our aim is not necessarily to find a set of parameters which all have a
perfectly linear effect on the power spectrum.  Rather, more
generally, we desire a set of parameters for which the power spectrum
can be approximated very accurately with a minimum of computational
effort. Then we can dispense with approximations of the likelihood
function, as in the Fisher matrix approach, and directly implement a
Monte Carlo parameter space search with a minimum of computational
effort.  This technique allows simple incorporation of prior
probabilities determined from other sources of data, and if fast
enough allows detailed exploration of potential systematic biases in
parameter values arising from numerous experimental and analysis
issues such as treatment of foregrounds, noise correlations, mapmaking
techniques, and model accuracy. 

We emphasize that with the accuracy of upcoming microwave background
power spectrum measurements, the cosmological parameters will be
determined precisely enough that their values will be significantly
affected by systematic errors from assumptions made in the analysis
pipeline and, potentially, numerical errors in evaluating theoretical
power spectra. The measurements themselves may also be dominated by
systematic errors. Extensive modelling of the impact of various
systematic errors on cosmological parameter determination will be 
{\it essential} before any parameter determination can be considered
reliable. This is the primary motivation for increasing the efficiency
of parameter error analysis. We discuss this point in more detail
in the last Section of the paper.

The following Section reviews physical processes affecting the cosmic
microwave background and defines our physical parameter set. The key
parameter, which sets the angular scale of the acoustic oscillations,
has been discussed previously in similar contexts
\cite{eis99,bon99,gri01} but has not been explicitly used as a
parameter. The other parameters describing the background cosmology
can then be chosen to model other specific physical effects.  Section
\ref{sec:power_spectra} displays how the power spectrum varies as each
parameter changes with the others held fixed.  The mapping between our
physical parameters and the conventional cosmological parameters
immediately reveals the structure of degeneracies in the cosmological
parameter space. Simple approximations for the effect of each physical
parameter are shown to be highly accurate for all but the largest
scales in the power spectrum.  We then determine the error region in
parameter space for an idealized model of the MAP experiment in
Sec.~\ref{sec:errors}, comparing with previous calculations, and test
the accuracy of our approximate power spectrum calculation. Finally,
Sec.~\ref{sec:discussion} discusses the potential speed of our method,
likelihood estimation, mapping between the sets of parameters,
accuracy, and systematic errors. The first appendix summarizes an
analysis pipeline going from a measured power spectrum to cosmological
parameter error estimates; the second appendix details several
numerical difficulties with using the CMBFAST code for the
calculations in this paper, along with suggested fixes.

Other recent attempts to speed up power spectrum computation
through approximations include the dASH numerical package \cite{kap02}
and interpolation schemes \cite{teg00}. Our method has the
advantage of conceptual simplicity and ease of use combined
with great speed and high accuracy.

\section{Cosmological Parameters and Physical Quantities}
\label{sec:parameters}

We consider the standard class of inflation-like cosmological
models, specified by five parameters determining the
background homogeneous spacetime 
(matter density $\Omega_{\rm mat}$, 
radiation density $\Omega_{\rm rad}$,
vacuum energy density $\Omega_\Lambda$, 
baryon density $\Omega_b$, and Hubble parameter $h$),
four parameters determining the spectrum of primordial perturbations
(scalar and tensor amplitudes $A_S$ and $A_T$ and power law
indices $n$ and $n_T$), and a single parameter $\tau$ describing
the total optical depth since reionization. We neglect additional
complications such as massive neutrinos or a varying vacuum equation
of state.
This generic class of simple cosmological models appears to be
a very good description of the Universe on large scales, and has the
virtue of arising naturally in inflationary cosmology.
Of course, any analysis of new data should first test whether this
class of models actually provides an acceptably good fit to
the data, and only then commence with determinations of cosmological
parameters.

What constraints do a measurement of the microwave
background power spectrum place on this parameter space?
A rough estimate assumes the likelihood of a particular set of
parameters is a quadratic function of the parameters (the usual Fisher
matrix technique); better is a Monte-Carlo exploration of parameter space 
in the region of a best-fit model. For either technique to give
reliable results, it is imperative to isolate the physical quantities
which affect the microwave background spectrum and understand their
relation to the cosmological parameters.  If a set of physical
quantities which have essentially orthogonal effects on the power
spectrum can be isolated, then extracting parameter space constraints
becomes much more reliable and efficient.

The five parameters describing the background cosmology
induce complex dependences in the microwave background power spectrum
through multiple physical effects. Using the physical energy densities
$\Omega_{\rm mat} h^2$, $\Omega_b h^2$, and $\Omega_\Lambda h^2$, as
has been done in numerous prior analyses, improves the situation 
but is still not ideal.  The characteristic physical scale in the power
spectrum is the angular scale of the first acoustic peak, 
so it is advantageous to use this scale as a
parameter which can be varied independently of any other parameters.
This angular scale is in turn determined by the ratio of the comoving
sound horizon at last scattering (which determines the physical
wavelength of the acoustic waves) to the angular diameter distance to
the surface of last scattering (which determines the apparent angular
size of this yardstick).  We first determine this quantity in terms of
the cosmological parameters, and then choose three other complementary
quantities.  The analytic theory underlying any such choice of
physical parameters has been worked out in detail
(see \cite{hs95a,hs95b,hs96,hw96}, and also \cite{wei01a,wei01b}).

The angular diameter distance to an object at a redshift $z$
is defined to be 
$D_A(z) =  R_0 S_k(r) / (1+z)$
where, following the notation of Peacock \cite{pea99}, $R_0$
is the (dimensionful) scale factor today and
\begin{equation}
S_k(r) = \cases{\sin{r},& $\Omega > 1$;\cr
                r,&  $\Omega = 1$;\cr
                \sinh{r},& $\Omega < 1$;}
\label{Skdef}
\end{equation}
the Friedmann equation provides the connection between
redshift $z$ and coordinate distance $r$. With a dimensionless
scale factor $a$ and a Hubble
parameter $H={\dot a}/a$, the differential relation is
\begin{equation}
H_0 R_0 dr = {- da \over \left[ (1-\Omega) a^2 + \Omega_\Lambda a^4
+ \Omega_{\rm mat} a + \Omega_{\rm rad}\right]^{1/2}}.
\label{drda}
\end{equation}
Thus the relation between scale factor and coordinate distance
becomes
\begin{eqnarray}
r &=& \left|\Omega-1\right|^{1/2}\nonumber\\
&\times&\int_a^1 {dx\over
\left[(1-\Omega)x^2 + \Omega_\Lambda x^4 + \Omega_{\rm mat} x + \Omega_{\rm
rad}\right] ^{1/2}}
\label{ra-integral}
\end{eqnarray}
and the angular diameter distance is just 
\begin{equation}
D_A(a) = aH_0^{-1}|\Omega -1|^{-1/2}S_k(r),
\label{daa}
\end{equation}
using the above expression for $r$. Here we have used the
relation $H_0R_0 = |\Omega -1|^{-1/2}$ which follows directly
from the Friedmann equation. 

The sound horizon is defined, in analogy to the particle
horizon, as
\begin{equation}
r_s(t) = \int_0^t {c_s(t') \over a(t')} dt'
\label{rsdef}
\end{equation}
where $c_s(t)$ is the sound speed of the baryon-photon fluid
at time $t$; to a very good approximation, before decoupling
the sound speed is given by \cite{hs95a}
\begin{equation}
c_s^2 = {1\over 3}\left(1+ {3\rho_b/4\rho_\gamma}\right)^{-1}.
\label{soundspeed}
\end{equation}
\begin{widetext}
Using this expression plus the equivalent expression to
Eq.~(\ref{drda}) connecting $dt$ and $da$ gives
\begin{equation}
r_s(a) = {1\over H_0\sqrt{3}}\int_0^a {dx \over
\left[\left(1 + {3\Omega_b\over 4\Omega_\gamma}x\right)
\left((1-\Omega)x^2 + \Omega_\Lambda x^4 + \Omega_{\rm mat} x + \Omega_{\rm
rad}\right)\right] ^{1/2}}.
\label{rsa-integral}
\end{equation}
\end{widetext}

The physical quantity relevant to the microwave background
power spectrum is
\begin{equation}
{\cal A}\equiv {r_s(a_*) \over D_A(a_*)}
\label{parameterA}
\end{equation}
where the two functions are given by Eqs.~(\ref{rsa-integral}) and
(\ref{daa}), and $a_*$ is the scale factor at decoupling. 
An accurate analytic fit for $a_*$ in terms of the cosmological
parameters has been given by Hu and Sugiyama \cite{hs96}:
\begin{eqnarray}
z_* &=& {1\over a_*} -1 = 1048
\left[1+0.00124(\Omega_bh^2)^{-0.738}\right]\nonumber\\
&&\qquad\times\left[1+ g_1(\Omega_{\rm mat}h^2)^{g_2}\right],\\
g_1 &\equiv& 0.0783(\Omega_b h^2)^{-0.238}
\left[1+39.5(\Omega_bh^2)^{0.763}\right]^{-1},\nonumber\\
g_2 &\equiv& 0.560\left[1+21.1(\Omega_bh^2)^{1.81}\right]^{-1}.\nonumber
\label{zstar}
\end{eqnarray}
This fit applies to standard thermodynamic recombination for a wide
range of cosmological models. (Note that while the formula was constructed
only for models with the standard value for $\Omega_{\rm rad}$, the
effect of a change of radiation density will come only through the
redshift of matter-radiation equality, and thus $\Omega_{\rm mat}h^2$
can be replaced by a factor proportional to 
$\Omega_{\rm mat}/\Omega_{\rm rad}$ to account for a variation
in the radiation density. The changes in $z_*$ are small anyway and have
little impact on our analysis.)
Since $a_* \ll 1$, the term proportional
to $\Omega_\Lambda$ in Eq.~(\ref{rsa-integral}) may be dropped and the
integral performed, giving \cite{hs95b}
\begin{eqnarray}
r_s(a) &=& {2\sqrt{3}\over 3}(\Omega_0 H_0^2)^{-1/2}
\left(a_{eq}\over R_{eq}\right)^{1/2}\nonumber\\
&&\qquad\qquad 
\times\ln{\sqrt{1+R}+\sqrt{R+R_{eq}}\over 1+\sqrt{R_{eq}}}.
\label{rsa-analytic}
\end{eqnarray}
where $R=3\rho_b/4\rho_\gamma$ is proportional to the scale factor.

If two cosmological models are considered which both have adiabatic
perturbations and the same value for ${\cal A}$, to high accuracy
their acoustic peaks
will differ only in height, not position. This is an obvious advantage
for constraining models. The peak positions, and thus ${\cal A}$, will
be extremely well constrained by power spectrum measurements.  Other
combinations which vary the peak heights only will be less well
determined.

Choosing four other parameters describing the background cosmology
must balance several considerations: (1) the new set of parameters
must cover a sufficiently large region of parameter space; (2) the
power spectrum should vary linearly or in some other simple way with
the new parameters; (3) the new parameters should be nearly
orthogonal in the cosmological parameter space; (4) the new parameters
should correspond to the most important independent physical effects
determining the power spectrum; (5) common theoretical prior
constraints, like flatness or standard radiation, should be simple to
implement by fixing a single parameter. No parameter set can satisfy
all of these constraints perfectly. We use the following parameters,
which provide a good balance between these criteria:
\begin{eqnarray}
{\cal B} &\equiv& \Omega_b h^2,\cr
{\cal V} &\equiv& \Omega_\Lambda h^2,\cr 
{\cal R} &\equiv& {a_*\Omega_{\rm mat} \over \Omega_{\rm rad}},\cr
{\cal M} &\equiv& \left(\Omega_{\rm mat}^2 + a_*^{-2}\Omega_{\rm rad}^2
\right)^{1/2} h^2.
\label{BVRMdef}
\end{eqnarray}
$\cal B$ is proportional to the baryon-photon density and thus
determines the baryon driving effect 
on the acoustic oscillations \cite{hs95a,hw96}.
$\cal R$ is the matter-radiation density ratio at recombination,
which determines the amount of early Integrated Sachs-Wolfe effect.
$\cal V$ determines the late-time Integrated Sachs-Wolfe effect arising
from a late vacuum-dominated phase, but otherwise represents a nearly
exact degeneracy (sometimes called the ``geometrical degeneracy''). 
$\cal M$ couples only to other small physical effects and
is an approximate degeneracy direction. This choice of parameters
is not unique, but this set largely satisfies the above criteria.

Given values for $\cal A$, $\cal B$, $\cal V$, $\cal R$, and $\cal M$,
they can be inverted to the corresponding cosmological parameters
by rewriting the definition of $\cal A$ in terms of $\cal B$,
$\cal V$, $\cal R$, $\cal M$, and $h$, then searching in $h$ until
the desired value for $\cal A$ is obtained. $\Omega_b$ and $\Omega_\Lambda$ 
then follow immediately, while $\Omega_{\rm mat}$ and $\Omega_{\rm rad}$
can be obtained with a few iterations to determine a precise value
for $a_*$. A less efficient but more straightforward method we have
implemented is simply
to search the 5-dimensional cosmological parameter space for the
correct values.

The other cosmological parameters which affect the microwave
background power spectrum, aside from tensor perturbations, are the
reionization redshift and the amplitude and power law index of the
scalar perturbation power spectrum. 
For reionization, we use the physical parameter 
\begin{equation}
{\cal Z}\equiv e^{-2\tau},
\label{Zdef}
\end{equation}
the factor by which the microwave background anisotropies at small
scales are damped due to Compton scattering by free electrons after the
Universe is reionized. At large scales, the temperature fluctuations
are suppressed by a smaller amount. 

The primordial perturbation amplitude cannot be measured directly, but
only the amplitude of the microwave background fluctuations. Some care
must be taken in the definition of the amplitude of the microwave
fluctuations so that the other physical parameters are not
significantly degenerate with a simple change in normalization. For
example, one common normalization is some weighted average of the
$C_l$'s over the smallest $l$ values, say $2<l<20$ \cite{bun97}.  This
is known as ``COBE-normalization'' because this is the range of scales
probed by COBE, which tightly constrains the normalization of 
the temperature fluctuations
at these scales \cite{gor94}. 
But although such a normalization is useful for
interpreting the COBE results, it is a bad normalization to adopt when
probing a much larger range in $l$-space. The reason is that the
lowest multipoles have a significant contribution from the Integrated
Sachs-Wolfe effect: if the total matter density changes significantly
between two models, the resulting low $l$ multipole moments will also
change. If the normalization is fixed at low $l$, then the result is
that the two models will be offset at high $l$ values. As a concrete
example, consider two models which differ only in ${\cal V}$, with
${\cal A}$, ${\cal B}$, ${\cal M}$ and ${\cal R}$ held fixed.  The
only physical difference between these two models is a difference in
the structure growth rate at low redshifts, and in particular the
fluctuations at the last scattering surface are identical. But if
these two models are COBE normalized, then every $C_l$ for $l=20$ is
offset between the two models.  Defining the normalization as an
average band power over a wider range of $l$ values \cite{bon99} is
better but not ideal. We advocate defining a normalization
parameter ${\cal S}$ by the amplitude of the perturbations on the scale
of the horizon at last scattering, corrected by the amount of
small-scale suppression in the $C_l$'s due to reionization. In
practice, for models with the same values of $n$, this
normalization essentially corresponds to the amplitude of the acoustic
oscillations at $l$ values higher than the first several peaks. 
Note that increasing ${\cal Z}$
while holding ${\cal S}$ and the other physical parameters fixed keeps
the microwave background power spectrum the same at high $l$ while
enhancing the signal at low $l$. 
The first few peaks cannot be used for normalization because they have
been subjected to driving by the gravitational potential as they cross
the horizon \cite{hw96}, so their amplitudes vary significantly with
${\cal B}$ and ${\cal R}$, as shown in the following Section. 

The primordial power spectrum of density perturbations is generally
taken to be a power law, $P(k) \propto k^n$, with $n=1$ corresponding
to the scale-invariant Harrison-Zeldovich spectrum.
Making the approximation that $k$ and $l$ have a direct
correspondence, the effect of $n$ on the microwave background
power spectrum can be modelled as
\begin{equation}
C_l(n) = C_l(n_0) \left(l\over l_0\right)^{n-n_0},
\label{n_effect}
\end{equation}
which is a good approximation for power law power spectra.  A
departure from a power law, characterized by $\alpha$ \cite{kos95}, can be
represented in a similar way.  Note that since the dependence is
exponential, a linear extrapolation is never a good approximation over
the entire interesting range $2<l<3000$. 
The choice of $l_0$ is arbitrary: changing $l_0$
simply gives a different overall normalization.

Finally, note that the amplitude of the tensor mode power spectrum
should be used as a separate parameter, not the common
choice of the tensor-scalar ratio. The tensor modes contribute
at comparatively large angular scales; they are significant
only for $l<100$ given current rough limits on the tensor
amplitude. Varying the tensor amplitude independently automatically
leaves the normalization parameter ${\cal S}$ fixed.

In the rest of the paper, we refer to the parameters
$\cal A$, $\cal B$, $\cal V$, $\cal R$, $\cal M$, $\cal S$, and $\cal Z$
as ``physical parameters'', as opposed to the usual ``cosmological
parameters'' $\Omega_b$, $\Omega_{\rm mat}$, $\Omega_{\rm rad}$,
$\Omega_{\rm vac}$, $h$, $Q^2$ (a quadrupole-based normalization),
and $z_r$.  

\section{Power Spectra}
\label{sec:power_spectra}

We now present the dependence of the temperature fluctuation power
spectrum on these physical parameters. 
As a fiducial model, we choose a standard cosmology with $\Omega =
0.99$, $\Lambda = 0.7$, $\Omega_{\rm mat} = 0.29$, $\Omega_b = 0.04$,
$h=0.7$, $n=1$, and full reionization at redshift $z_r = 7$, 
with standard neutrinos. We compute power spectra for the fiducial
model and for evaluating the various numerical derivatives using
Seljak and Zaldarriaga's CMBFAST code \cite{sel96,zal98,zal00}.
Note that the model is slightly open
rather than flat to compensate for numerical instabilities in the CMBFAST code;
see the discussion in the Appendix. The corresponding physical
parameters are ${\cal A}_0=0.0106$, ${\cal B}_0=0.0196$, 
${\cal V}_0=0.338$, ${\cal M}_0=0.154$, and ${\cal R}_0 = 3.252$.
Since the dependence of the spectrum on ${\cal S}$ is trivial, its definition
is arbitrary; it can be taken, for example, as the amplitude of a
particular high peak, or as the equivalent scalar perturbation
amplitude $A_S$ at the scale which the power law $n$ is defined for a
model with no reionization.  We do not consider tensor perturbations
in this paper, which can be analyzed separately by the same
techniques; the tensor perturbations are simpler because they depend
on fewer parameters than the scalar perturbations. 
None of the following results depend qualitatively on the
fiducial model chosen.

\begin{figure*}
\includegraphics{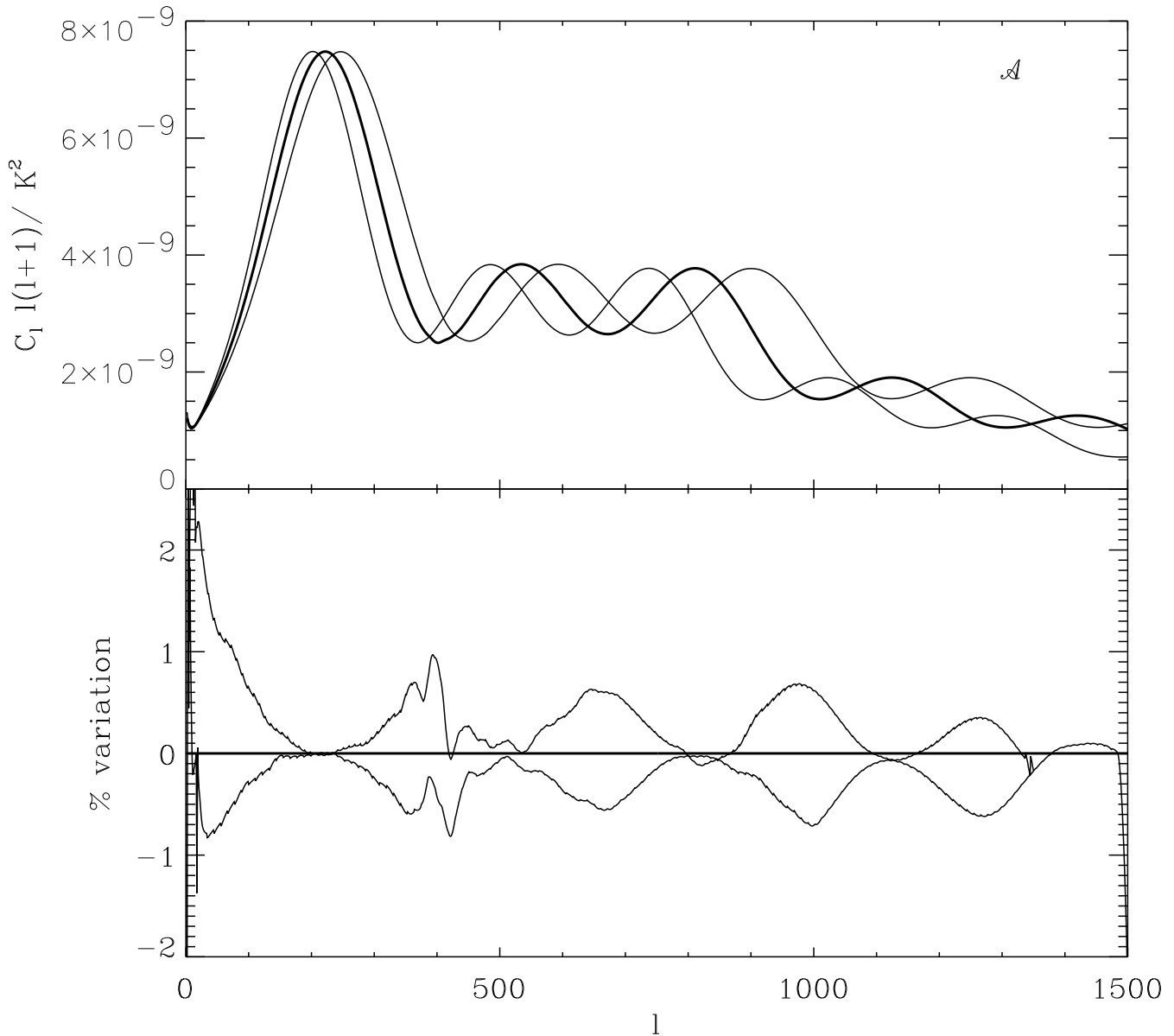}
\caption{Temperature power spectra for the fiducial cosmological
model (top panel, heavy line), 
plus models varying the parameter ${\cal A}$ upwards by
10\% (curve shifts to lower $l$ values) and downwards by 
10\% (curve shifts to higher $l$ values) while keeping
${\cal B}$, ${\cal M}$, ${\cal V}$, ${\cal R}$, ${\cal S}$ and $n$ fixed.
The bottom panel displays the fractional error between the fiducial
model and the other two models with $l$-axes rescaled by the factor
${\cal A}/{\cal A_0}$.}
\label{Cl_A}
\end{figure*}

The top panel of
Fig.~\ref{Cl_A} displays power spectra as the parameter ${\cal A}$ varies
while the other physical parameters are held fixed.  It is clear that
the effect of ${\cal A}$ is {\it only} to determine the overall angular
scale, except for the small Integrated Sachs-Wolfe effect for $l<200$. 
For the bottom panel,  the $l$-axis of each power spectrum is rescaled
to give $C_l$ as a function of $l{\cal A}/{\cal A}_0$, 
and then the percent difference
between the rescaled model and the fiducial model is plotted. 
The small variation of
the scaled power spectrum between around $l=50$ and $l=200$ is due to
the early ISW effect and can be accurately modelled as a linear
variation, while between $l=2$ and $l=50$ the variation with ${\cal A}$ is
more complicated due to the late-time ISW contribution.  For $l > 200$,
the scaled power spectra match to within 1\%; 
the residual discrepancy is likely due to numerical
inaccuracy. Since ${\cal A}$ will be tightly constrained by the peak
positions, this numerical error will not affect the determination of
${\cal A}$, but may contribute a systematic error to the determination of
other parameters at the 1\% level.

\begin{figure*}
\includegraphics[width=\textwidth]{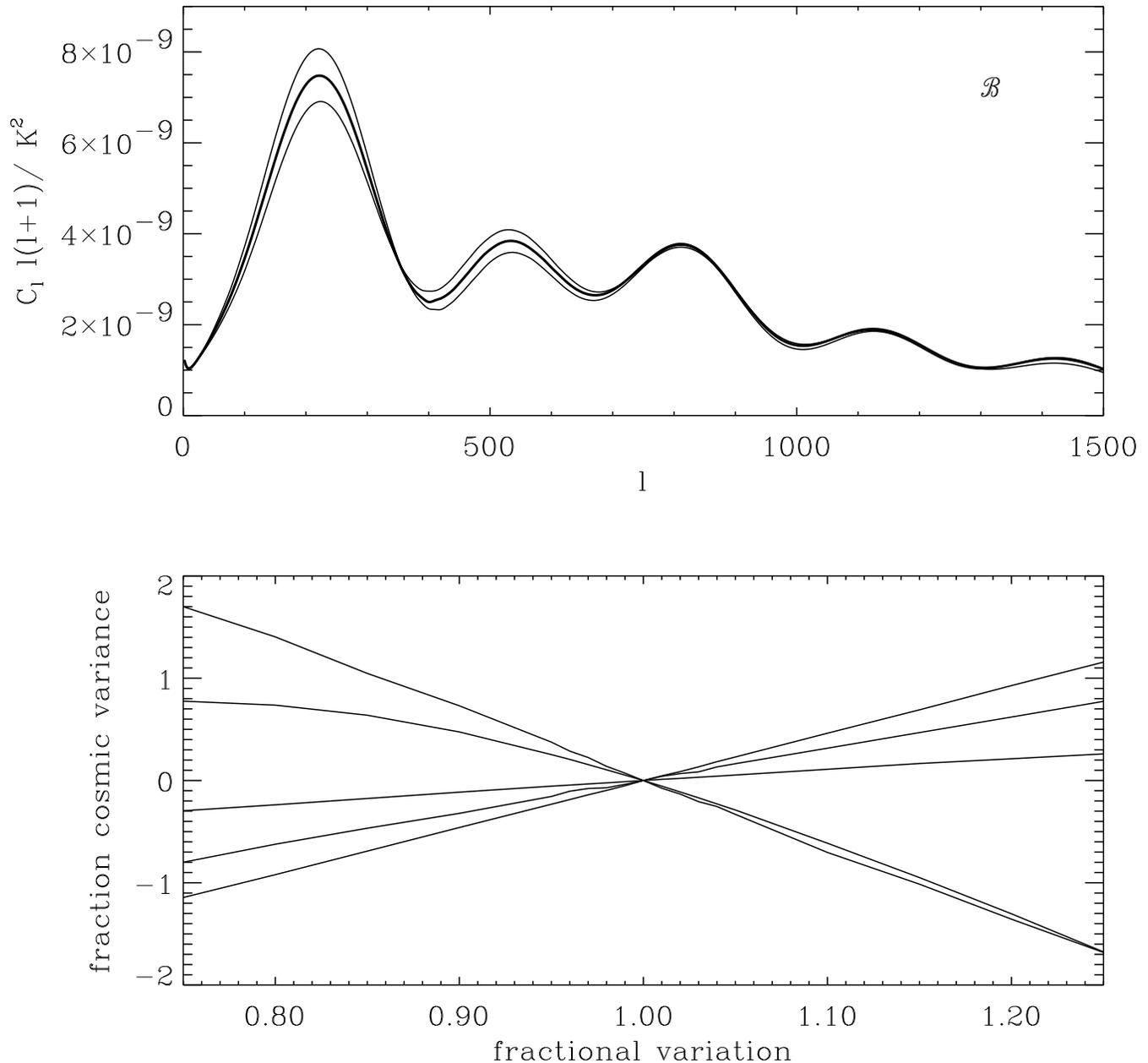}
\caption{Temperature power spectra for the fiducial cosmological model
(top panel, heavy line), plus models varying the parameter ${\cal B}$
upwards by 25\% (higher first peak) and downwards by 25\% (lower first
peak) while keeping ${\cal A}$, ${\cal M}$, ${\cal V}$, ${\cal R}$,
${\cal S}$, and $n$ fixed.  The bottom panel displays $C_l$ as a function of
${\cal B}$, for $l=50$, 100, 200, 500, and 1000. The horizontal axis
shows the fractional change in $\cal B$, while the vertical axis gives
the change in $C_l$ as a fraction of the cosmic variance at that
multipole.}
\label{Cl_B}
\end{figure*}

\begin{figure*}
\includegraphics[width=\textwidth]{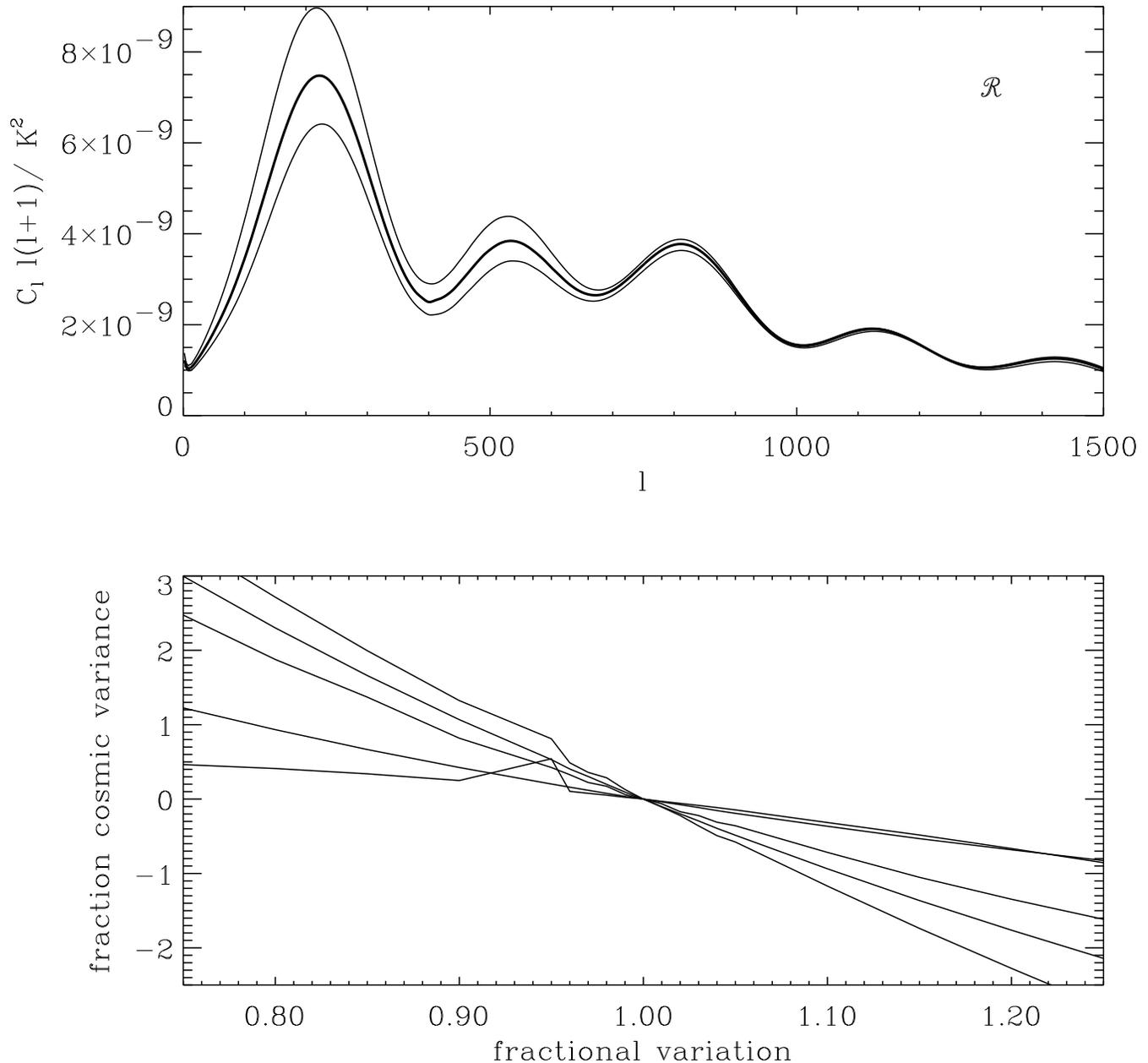}
\caption{Same as Fig.~\ref{Cl_B}, except varying the parameter
${\cal R}$ while keeping the others
fixed. Larger ${\cal R}$ increases the peak heights. The substantial
glitch in the bottom panel is due to systematic errors in CMBFAST.}
\label{Cl_R}
\end{figure*}

\begin{figure*}
\includegraphics[width=\textwidth]{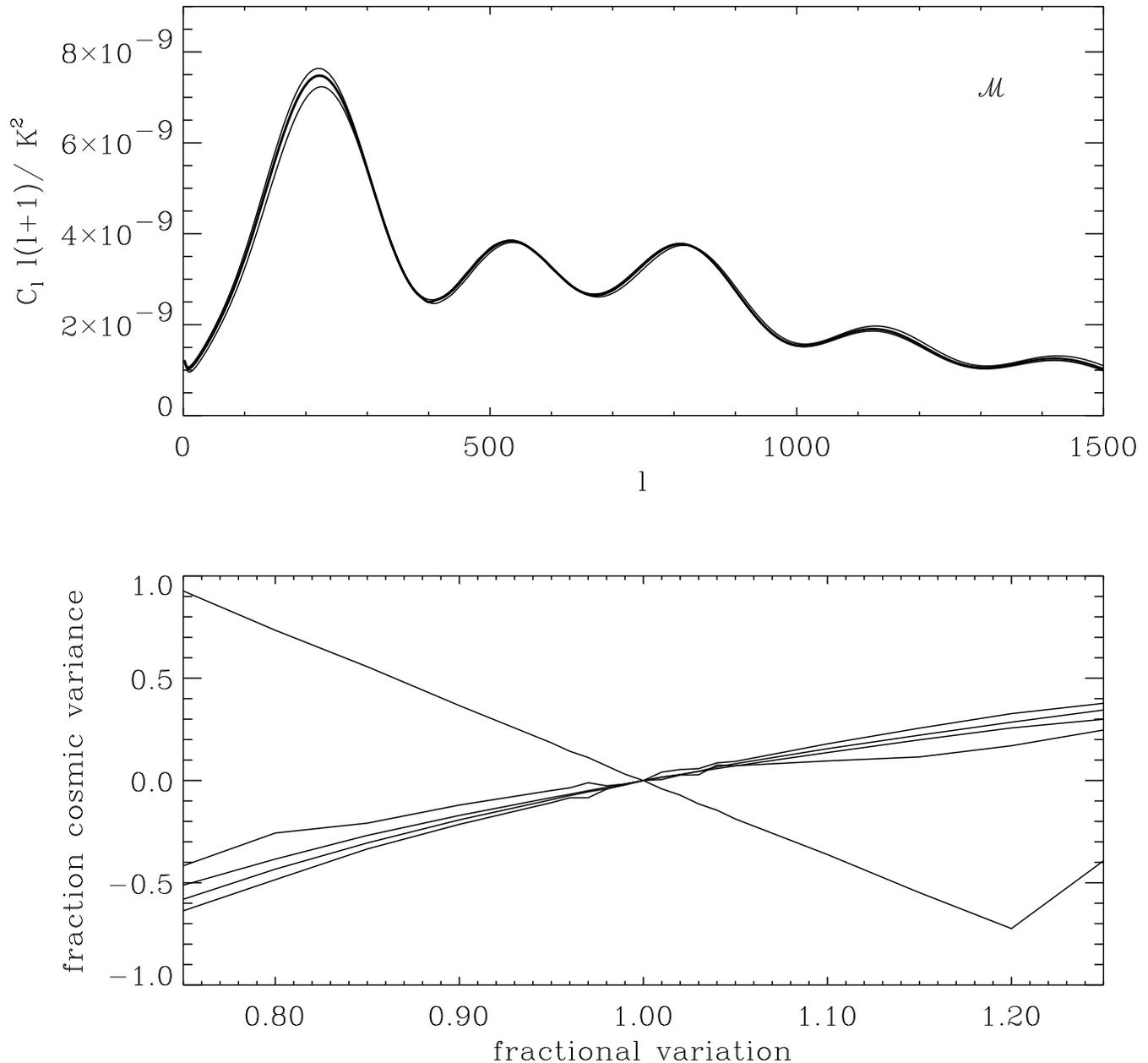}
\caption{Same as Fig.~\ref{Cl_B}, except varying the parameter
${\cal M}$ while keeping the others
fixed. Larger $\cal M$ increases the first peak height. Again, the glitch in
the bottom panel is due to systematic errors in CMBFAST.}
\label{Cl_M}
\end{figure*}

\begin{figure*}
\includegraphics[width=\textwidth]{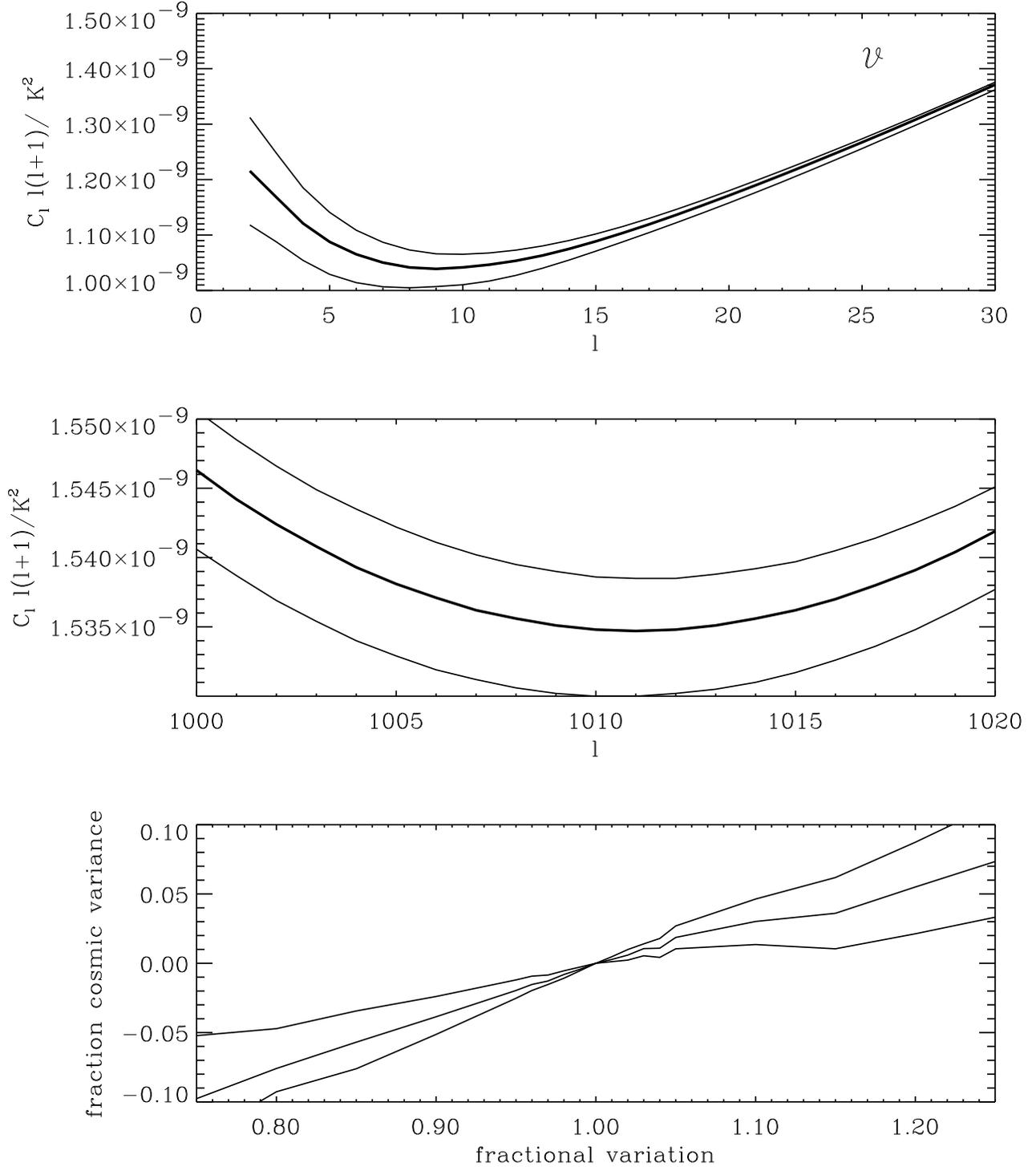}
\caption{Varying the the parameter ${\cal V}$, keeping the others
fixed, for low $l$ (top) and higher $l$ (center) values; note
the scales of the vertical axes.  Only the lowest multipoles
have any significant variation, arising from the Integrated
Sachs-Wolfe effect at late times; the bottom panel shows the dependence
of the $l=5$, 10, and 20 multipoles on ${\cal V}$. A linear 
approximation is rough but reasonable; higher-order approximations
will model this dependence better. At higher $l$, the variation
between the curves is a measure of the numerical accuracy of the code
generating the $C_l$ curves.}
\label{Cl_V}
\end{figure*}

\begin{figure*}
\includegraphics[width=\textwidth]{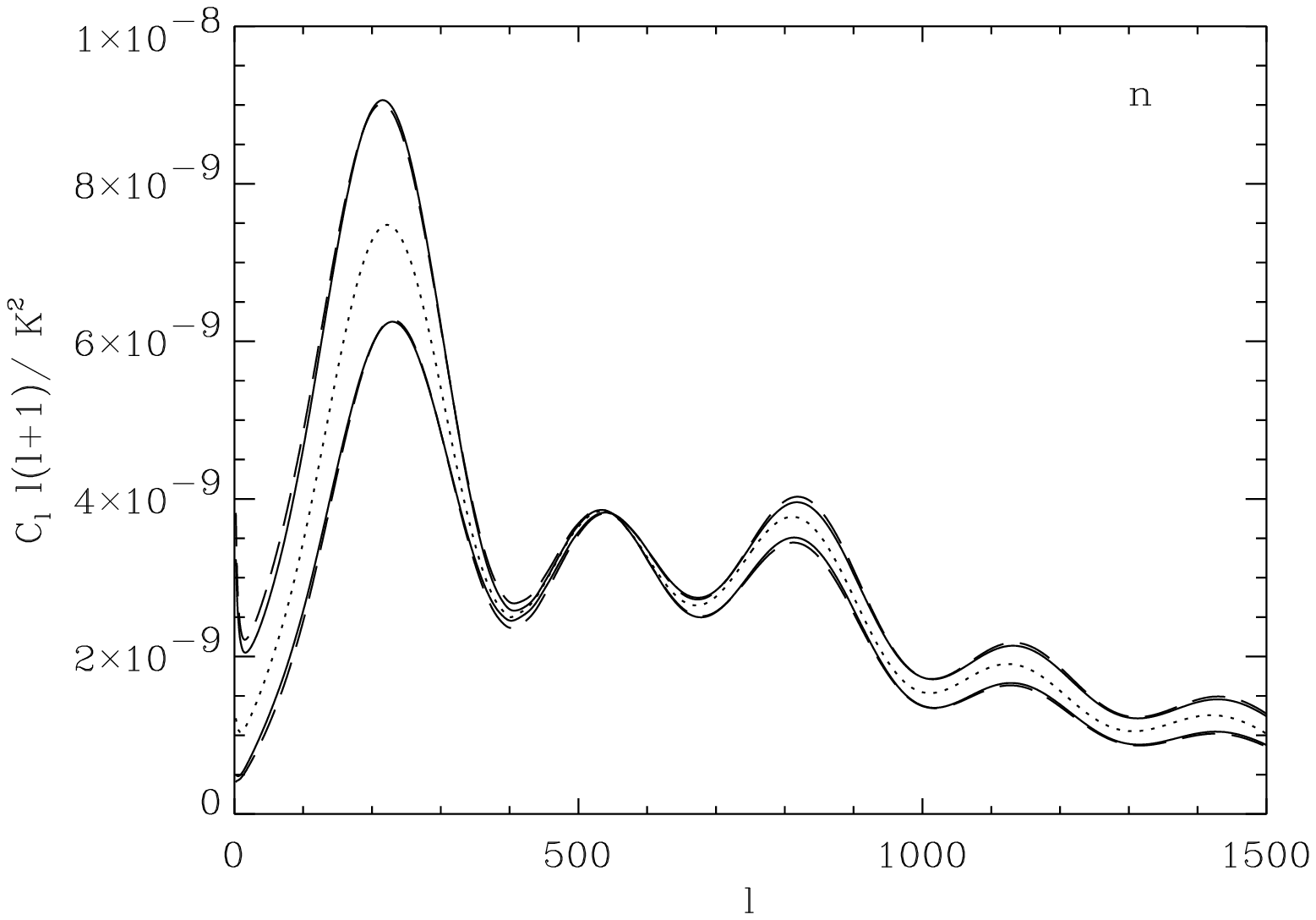}
\caption{A comparison of computed (solid) and approximated (dashed) 
power spectra for
different values of the scalar index $n$, for $n=0.8$ and
$n=1.2$ (dashed), where the approximated values have been calulated by
applying Eq.~(\ref{n_effect}) to the $n=1$ model (dotted). Note the
displayed variation in $n$ is roughly three times larger than the
constraint MAP will impose (see Fig.~\ref{errors_physical_full}).}
\label{Cl_n}
\end{figure*}

Figure \ref{Cl_B} displays the power spectrum variation with ${\cal B}$,
keeping the other parameters fixed. This plot clearly shows the
well-known baryon signature of alternating peak enhancement and
suppression. The variation of a particular multipole $C_l$ with ${\cal B}$ is
highly linear for $l>50$ for variations of up to 20\%; 
for all multipoles, the dependence is linear
for variations of up to 10\%. 
A figure of merit for a linear approximation
to the power spectrum is whether the errors
in the approximation are negligible compared to the cosmic statistical
error at a given multipole. The bottom panel of Fig.~\ref{Cl_B}
shows just how good the linear approximation is for several
selected $l$ values. The horizontal
axis plots the fractional change
in the parameter $\cal B$.   
The vertical axis plots the ratio of the
change in $C_l$, $C_l(B) - C_l(B_0)$, to the statistical error from
the cosmic variance at that multipole, $\sigma_l =
C_l(B_0)\sqrt(2/(2l+1))$. 
(Any measurement of the multipole $C_l$ is
subject to a statistical error at least as large as this cosmic
variance error, due to the finite number of modes sampled on the sky.)
The linear dependence remains valid even
near the ``pivot'' points between a peak and a trough. Thus a linear
extrapolation using computed numerical derivatives is highly accurate
at reproducing the ${\cal B}$ dependence; the error in such
an approximation appears to be dominated by systematic numerical
errors in computing the power spectrum, and funny behavior of
some of the multipoles for small variations in $\cal B$ is surely
due to systematic errors. 

Figure \ref{Cl_R}
shows the variation of the power spectrum with respect to the
parameter ${\cal R}$, keeping the others fixed, with the
corresponding dependence of specific multipoles. For smaller values of
${\cal R}$, matter-radiation equality occurs later, so the
Universe is less accurately described as matter-dominated at
the time of last scattering, leading to
an increased amplitude of the first few peaks from the Integrated
Sachs-Wolfe effect at early times.  
The dependence on ${\cal R}$ is not quite as
linear as for ${\cal B}$, but still the spectrum varies highly
linearly with ${\cal R}$ for parameter variations of 10\% for
$l>30$. Figure \ref{Cl_M} shows the same for the parameter $\cal M$,
which is an approximate degeneracy direction: the power spectrum
changes noticably only in the neighborhood of the first peak.
If the standard three massless neutrino species is imposed as
a prior, $\cal M$ is fixed.

Variation of $\cal V$ holding the others fixed is nearly
an exact physical degeneracy \cite{bon94,bon99}, broken only by the
late-time ISW effect at low multipole moments displayed
in Fig.~\ref{Cl_V}. This variation at small $l$ values
is not well approximated by simple linear extrapolation; accurate
approximation must rely on more sophisticated schemes, but the
total contribution to constraining cosmological models will
have comparatively little weight due to the large cosmic variance. At higher
$l$ values, the power spectra 
should be precisely degenerate, and the extent to which they are not
is a measure of numerical inaccuracies. Note that the inaccuracies are
in the form of systematic offsets at levels smaller than a
percent. When analyzing data, the variation in ${\cal V}$ can simply
be set to zero for all $l$ values higher than the first acoustic
peak. If a flat universe is imposed as a prior
condition, the parameter ${\cal V}$ is then fixed.

Aside from initial conditions, reionization is the other physical
effect which can have a significant impact on the power
spectrum. Note that
the effective normalization we use depends on ${\cal Z}$, so that as ${\cal Z}$
varies, the power spectrum retains the same amplitude at high $l$.  As
discussed in the previous section, for large $l$ the
effect of changing the total optical depth is effectively just a
change in the overall amplitude. Thus for our set of parameters,
variations in ${\cal Z}$ while holding the other parameters fixed change the
power spectrum only at the largest scales. For this reason, we do not
consider reionization further here.
Note that for polarization, reionization will
generate a new peak in the polarization power spectrum at low $l$, and
this effect may require more complicated analytic modelling than a
simple linear extrapolation \cite{zal97b}.

Finally, we show the effect of changing the spectral index
$n$ of the primordial power spectrum in Fig.~\ref{Cl_n}. 
Equation (\ref{n_effect}) gives a very good approximation
to the effect of changing $n$; the largest errors
are at small $l$ and in the immediate region of the
first few peaks and troughs, amounting to less than 10\% of the cosmic
variance for any individual multipole. 

We have now displayed simple numerical or analytic approximations
for the effect of all parameters with a sizeable impact on
the cosmic microwave background power spectrum, for $l>50$ and
for parameter variations which are in the range to which the
parameters will be restricted by upcoming microwave background
maps. For a model which differs from a fiducial model in several
different parameters, it is in principle ambiguous in what order
to do the various extrapolations and approximations. For example,
should a reionization correction be applied before or after
the ${\cal A}$ parameter extrapolation? In practice, however, most of the
parameter ranges involved will be quite small, particularly
for the ${\cal A}$ parameter, and the order in which all of the
approximations are applied to the fiducial spectrum is essentially
irrelevant. We test the validity of the various approximations
for arbitrary models below. 

We also note that while we have demonstrated explicitly the
validity of linear extrapolations and other simple approximations
in computing deviations from a particular fiducial power spectrum,
we have not shown explicitly that this is true for any fiducial
cosmological model. But current measurements point towards a
Universe well-described by a model close to the fiducial one
considered here, and in the unlikely event that the actual
cosmology is significantly different, the various approximations
are simple to check explicitly for any underlying model.

Some of the linear extrapolations lose accuracy for the smallest $l$
values.  It is not clear whether any simple approximation is
sufficient for reproducing the power spectrum for small $l$ values,
because a variety of different effects contribute in varying
proportions. While these large scales have significant cosmic variance
and do not have very much statistical weight compared to the rest of
the spectrum, several parameters ($\cal V$, $\cal Z$, $\cal T$, $n_T$)
have significant effects {\it only} at small $l$, and accurate limits
on these parameters will require approximating these moments.  One
possible approach is to use quadratic or higher-order extrapolations.
We have not explored the efficacy of such approximations, and they may
require a more accurate numerical code, but promise to be a good
solution. A more complicated approach
would be to develop analytic fitting formulas.  Steps in this
direction have been taken by Durrer and collaborators \cite{dna01},
but their solutions rapidly lose accuracy for $l>10$, and a variety of
physical effects are neglected which contribute at the few percent
level or more.  Another possibility is an expanded numerical
approximation, where the effects of our parameters are fit as linear
or quadratic, but instead of fitting the total $C_l$ dependence, the
various separate terms (i.e., quadratic combinations of Sachs-Wolfe,
Integrated Sachs-Wolfe, doppler, and acoustic effects) contributing to
each $C_l$ are modelled independently.
In the estimates below, we will not consider tensor perturbations and
fix the parameter $\cal Z$, while $\cal V$ is a degenerate direction;
we thus sidestep the issue of approximating the power spectrum for
low $l$, and do not discuss the issue further here.

\section{Error Regions for Cosmological Parameters}
\label{sec:errors}

Given some data, we want to determine the error region in parameter
space corresponding to some certain confidence level in fitting the
data. This must be done by looking around in parameter space in the
vicinity of the best-fitting model. With a fast computation of the
theoretical power spectrum for different points in parameter space in
hand, a straightforward Metropolis algorithm can be used to construct
a Monte-Carlo exploration of the parameter space.  Refined techniques
with the label of Markov Chains have recently been applied to the
microwave background \cite{chr01,lew02}. Our approximate power spectrum
evaluation renders these techniques highly efficient.

\begin{figure*}
\includegraphics[width=\textwidth]{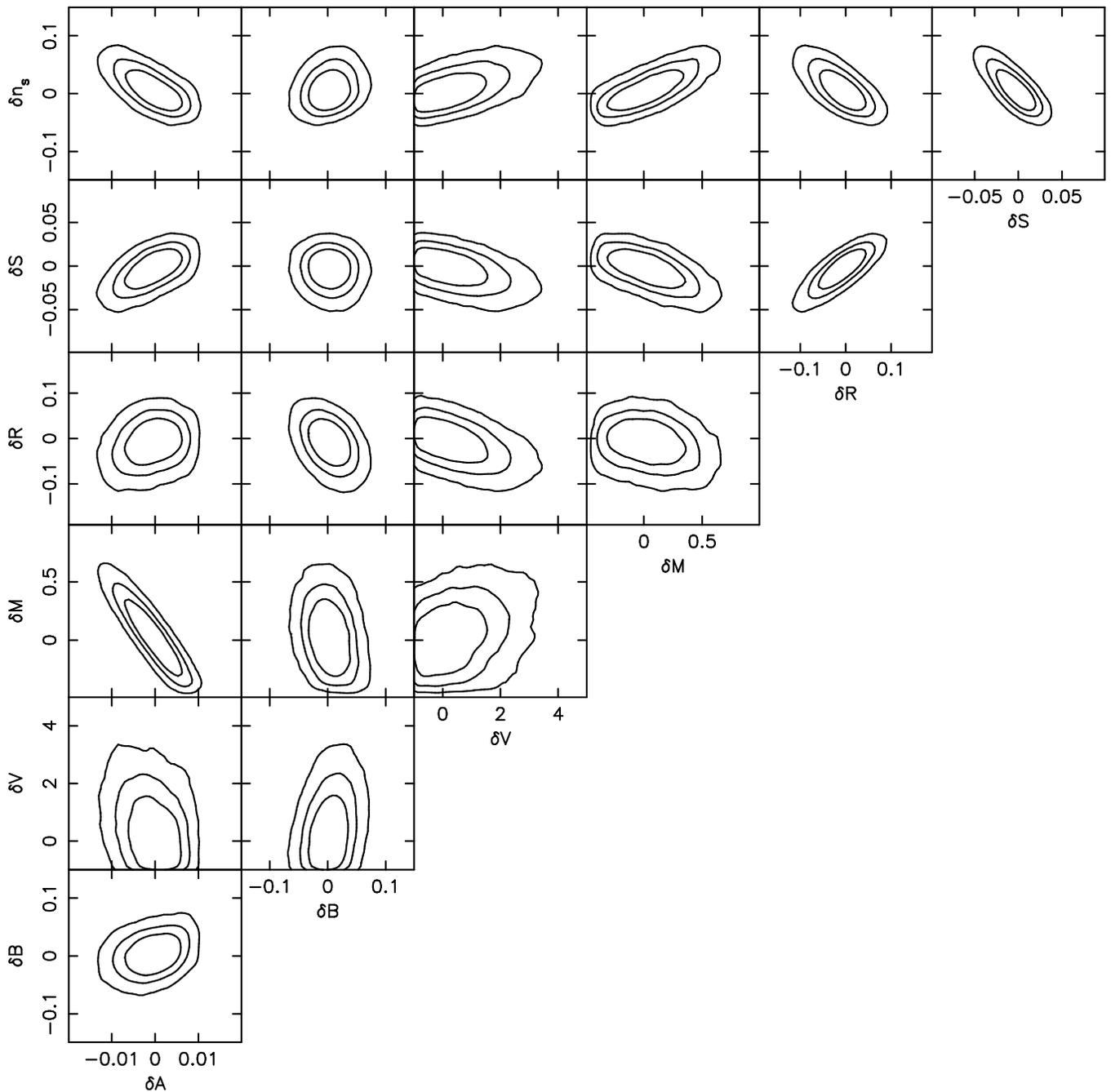}
\caption{Parameter error contours anticipated for the MAP satellite in
  the physical parameter space.  In each panel, the entire likelihood region
  has been projected onto a particular plane. The axis refer to the fractional
  variation for each parameter with respect to the fiducial one, e.g. $\delta
  A = (A-A_{\rm fid})/A_{\rm fid}$.  The contours show 68\%, 95\%, and 99\%
  confidence levels.  }
\label{errors_physical_full}
\end{figure*}

\begin{figure*}
\includegraphics[width=\textwidth]{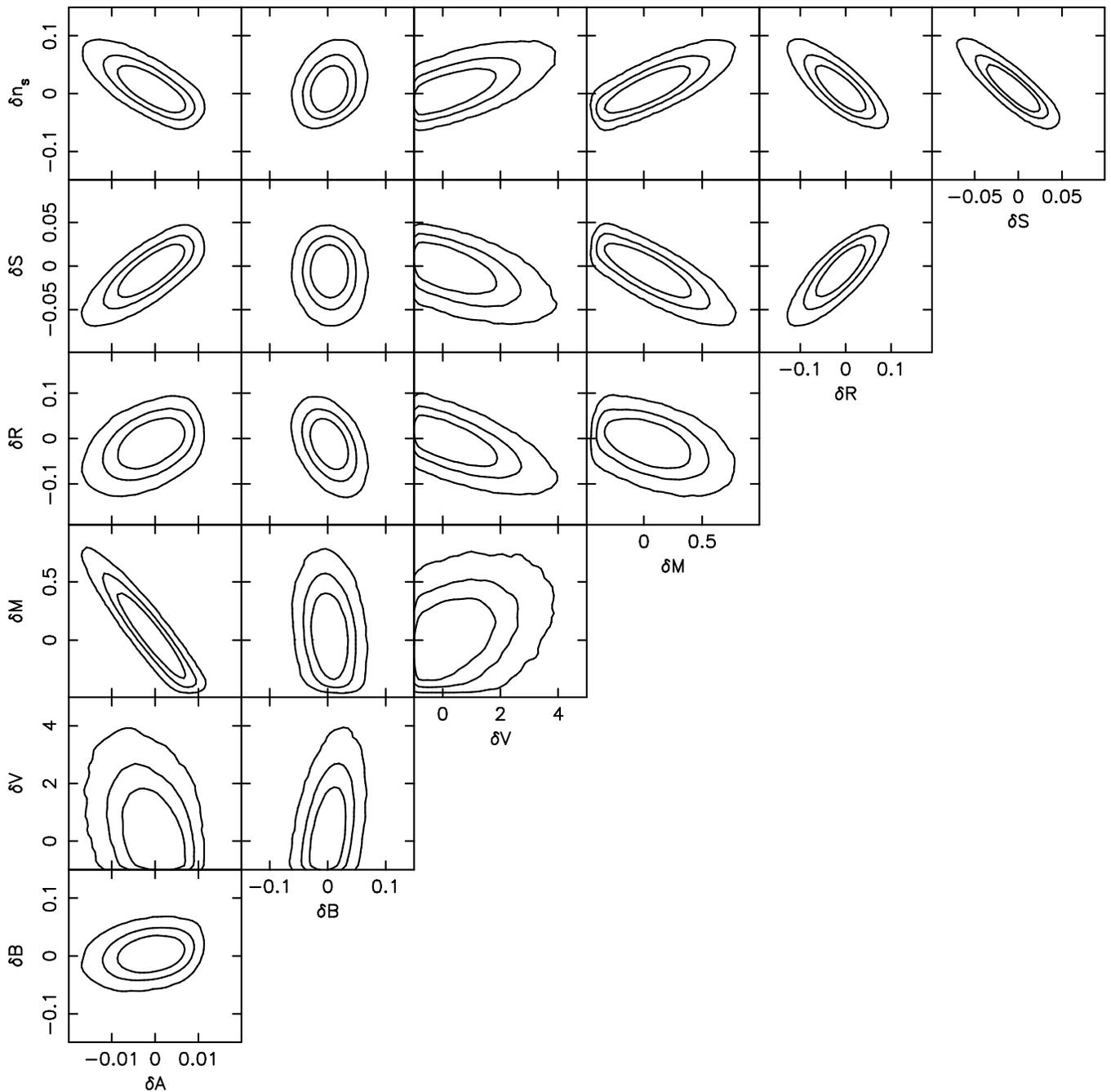}
\caption{The same as Fig.~\ref{errors_physical_full}, except using
the Fisher matrix approximation to the actual likelihood. The contours
in the two cases are very similar.}
\label{errors_physical_fisher}
\end{figure*}

\begin{figure*}
\includegraphics[width=\textwidth]{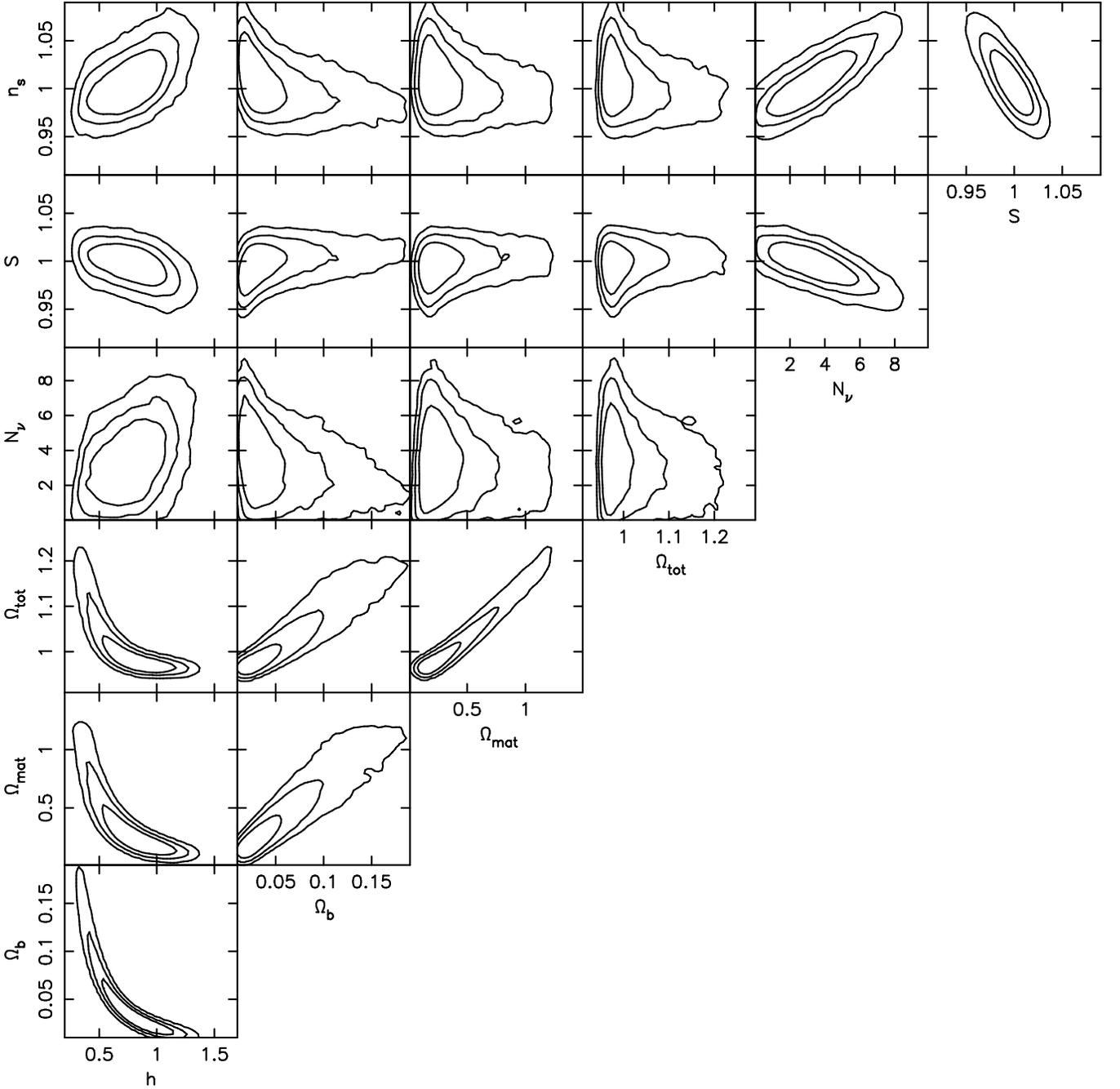}
\caption{The likelihood distribution in Fig.~\ref{errors_physical_full}
mapped into the cosmological parameter space.}
\label{errors_cosmological_full}
\end{figure*}

\begin{figure*}
\includegraphics[width=\textwidth]{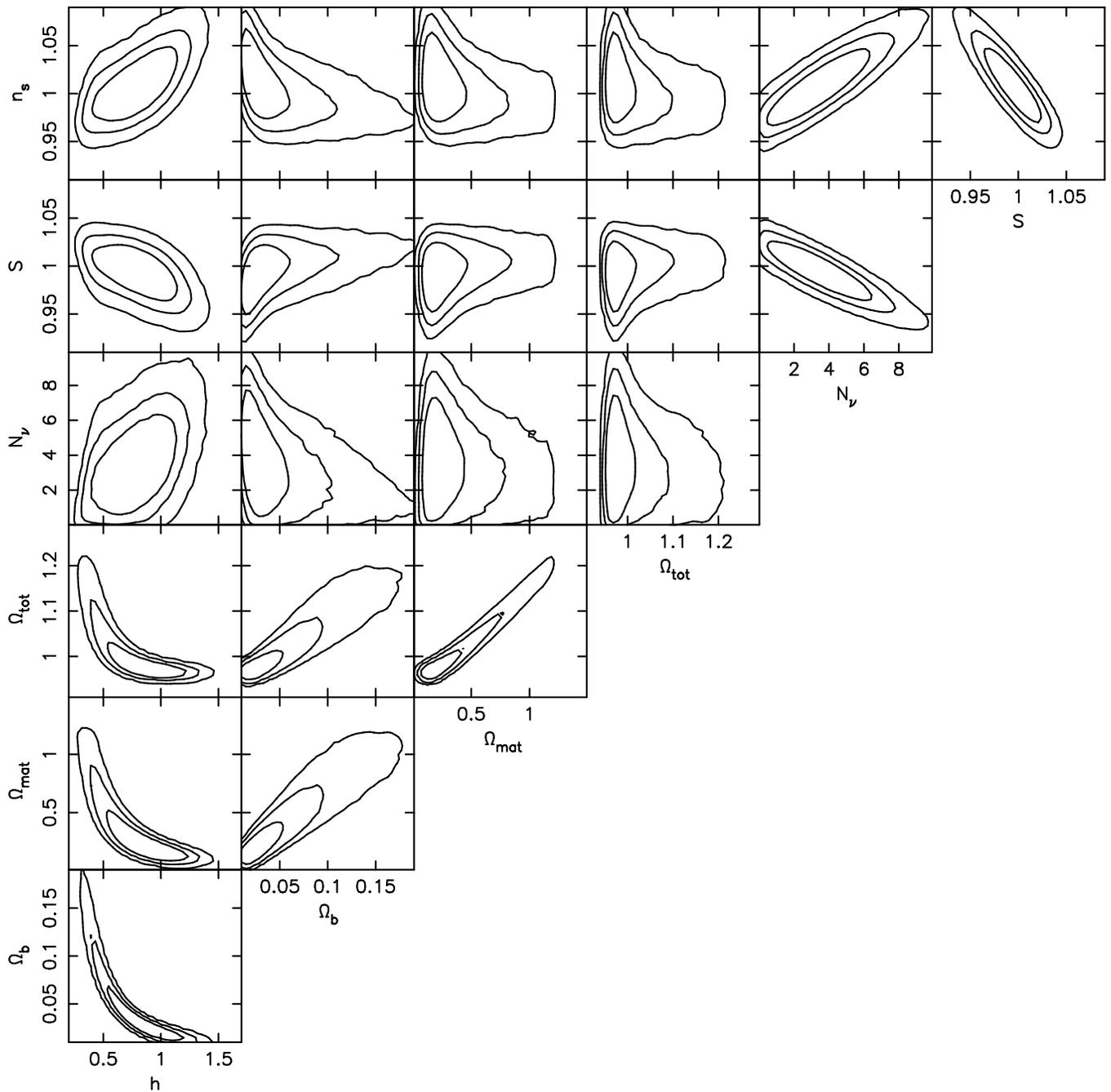}
\caption{The likelihood distribution in Fig.~\ref{errors_physical_fisher}
mapped into the cosmological parameter space. These contours, derived
from a Fisher matrix approximation, are very close
to those in Fig.~\ref{errors_cosmological_full}.}
\label{errors_cosmological_fisher}
\end{figure*}

As a demonstration, we consider estimated power spectrum measurement errors
drawn from the $\chi^2_{2l+1}$
probability distribution derived in Ref.~\cite{kno95}. We assume
that the Universe is actually described by a particular fiducial
cosmological model. With the
simple assumption that the measurement errors for each 
$C_l$ are uncorrelated, the
likelihood of a cosmological model being consistent with a measurement
of the fiducial model becomes trivial to compute.
Then we map the likelihood around the fiducial model via a Markov
Chain of points in parameter space, starting with the fiducial model
and moving to a succession of random points in the space using a
simple Metropolis-Hastings algorithm.  Sophisticated convergence tests
can be applied to the chain to determine when the distribution of
models in the Monte Carlo has converged to a representation of the
underlying likelihood function. The chain of models should be
constructed in the physical parameter space: the roughly orthogonal
effects of the physical parameters on the power spectrum leads to an
efficient Monte Carlo sampling of the space. For each model in the
chain, the cosmological parameters of the model are computed and
stored along with the physical parameters.

The commonly-used Fisher matrix approximation \cite{jun96} expands the
likelihood function in a Taylor series around the most likely model
and retains only the quadratic term. With Gaussian uncorrelated
statistical errors, the Fisher matrix approximation becomes exact
whenever the parameter dependence of the $C_l$'s is exactly linear in
all parameters. Since the power spectrum is nearly linear in most of
our physical parameter set, we can check our Monte Carlo results by
comparing with the Fisher matrix likelihood computed with the physical
variables. 

As a simple illustration, we determine the error region
corresponding to a simplified model of the power spectrum
which will be obtained by the MAP satellite, currently
collecting data. MAP's highest frequency channel has a 
gaussian beam with a full-width-half-max size of about 0.21 degrees.
MAP will produce a full-sky map of the microwave background
at this resolution with a sensitivity of around 35 $\mu$K
per 0.3 square-degree pixel. We neglect eventual sky cuts
and assume all $C_l$ estimates are uncorrelated.
We consider models with only scalar perturbations
and fixed reionization. The physical parameter space is thus
${\cal A}$, ${\cal B}$, ${\cal M}$, ${\cal R}$, ${\cal V}$ and $n$. The other
parameters we have fixed (tensor perturbations, ${\cal Z}$)
affect the power spectrum only at small $l$,
and will have little effect on the physical parameters considered.
We assume the underlying cosmology is described by the
fiducial model of the previous section. Fundamental physical constraints
have been enforced: $\Omega_b < \Omega_{\rm mat}$ and
all densities must be positive. Additionally, we only consider
models with $\Omega_{\rm vac} < 2$ and discard any models which
are not invertible from the physical to the cosmological
parameters (these amount to a handful of models with extreme
parameter values).

Figure \ref{errors_physical_full} shows the anticipated MAP error
contours in the physical parameter space, extracted from a Markov
Chain of $3\times 10^4$ models. 
The full 7-dimensional likelihood region is
projected onto all pairs of parameters.  The parameter ${\cal V}$ is
essentially unconstrained by MAP, while ${\cal M}$ has a 1-$\sigma$
error of around 30\%. Measurements extended to smaller scales will not
significantly improve constraints on these two parameters: they
represent true physical degeneracies in the model space. The other
parameters are well constrained.  Approximate 1-$\sigma$ errors are:
0.5\% for $\cal A$, 5\% for $\cal R$, 3\% for $\cal B$, 2\% for $\cal
S$, and 3\% for $n$.  The error regions are elliptical, because the
likelihood region in these parameters is well-approximated by a
quadratic form. The residual correlations involving $\cal S$ and $n$
will be lifted by measurements out to higher $l$ values; the
parameters are largely uncorrelated in their power spectrum effects.
Figure \ref{errors_physical_fisher} shows a Fisher matrix estimate of
the same likelihood. The two sets of likelihood contours are very
close, as expected since we have shown explicitly that the power
spectrum varies nearly linearly in the physical variables over a
parameter region consistent with MAP-quality data.  The most
notable exception is the variable $\cal A$, which does not give a
linear variation, but it is so well constrained that it is effectively
fixed when considering its impact on determining the other
parameters. The contours will also differ slightly due to the
nonlinear behavior of $n$, but the general excellent agreement
demonstrates that our Monte Carlo technique works as expected.

In Fig.~\ref{errors_cosmological_full}, we replot
Fig.~\ref{errors_physical_full} in the cosmological parameter
space.  This procedure is trivial, since we calculate the cosmological
parameter set for each model. The resulting contours are deformed,
reflecting the non-linear relation between the physical and
cosmological parameter sets. Clearly, a Fisher matrix approximation in
the cosmological variables, as has been done in numerous analyses,
will give significantly incorrect error regions \cite{bon99}. 
Note, however, that for MAP-quality data,
the Fisher matrix errors in the physical variables
projected into the cosmological variables gives a highly accurate
result, Fig.~\ref{errors_cosmological_fisher}, in contrast to the
claims in Ref.~\cite{bon99}. The Fisher matrix approximation in the
physical parameter space can be used to further increase the efficiency
of error region evaluation, which may be useful in the cases where
the compuation of the error regions is dominated by evaluation of 
the likelihood for a given model from actual data. If the likelihood
evaluation is at least as efficient as the power spectrum evaluation,
then the increased efficiency from a Fisher matrix approximation
may not result in dramatic increases in computational speed, since
in both cases the physical parameter space must be sampled and parameter
conversion performed at each point to construct the likelihood contours
in the space of the cosmological parameters. 

It is useful to plot error regions in both sets of parameters.
Physical parameters represent the actual physical
quantities which are being directly constrained by
the measurements, and the error regions are nearly
elliptical. The cosmological parameters
are useful theoretically and for comparisons with
other data sources, i.e. large-scale structure
or supernova standard candles. One advantage of determining
the error region via a Monte Carlo technique is that
prior constraints on cosmological parameters can be
implemented simply.

When the Fisher matrix approximation is computed in the physical
parameter space, it is highly accurate. This does not actually
help much, however: if information about cosmological
parameters is desired, transforming from the physical to the
cosmological parameter space requires some kind of sampling
of the likelihood region. The most efficient way to do this is
via Monte Carlo, so the additional computational overhead
in performing a Monte Carlo likelihood evaluation of the
physical variable likelihood instead of a Fisher matrix
approximation is negligible. 

\begin{figure}
\includegraphics[width=3.4in]{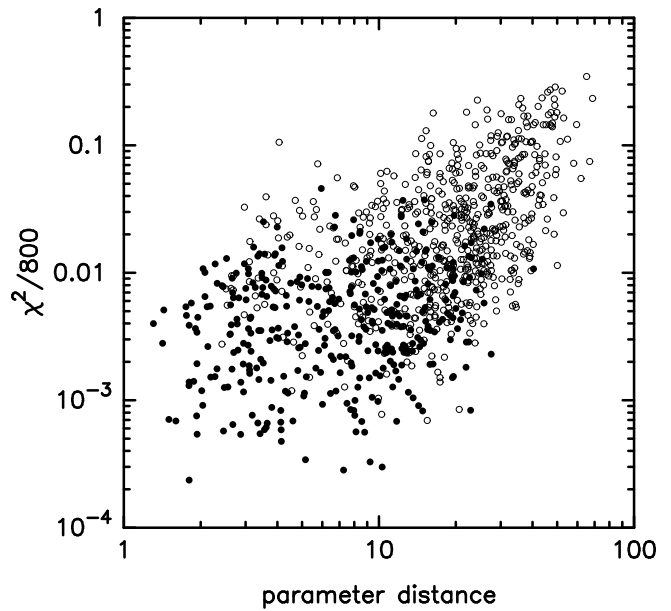} 
\caption{A plot of the difference between approximate
and numerical power spectra versus distance from the
fiducial model in parameter space, 
for 1000 cosmological models
drawn randomly from the distribution in Fig.~\ref{errors_physical_full}.
The solid circles are for models with $2<N_\nu<4$.}
\label{error_plot}
\end{figure}

To test the validity of our numerical power spectrum approximations,
we choose 1000 models at random from the above Monte Carlo set of
cosmological models.  We compute each numerically using CMBFAST, and
then compare the computed and approximated models. The two are
compared using the statistic
\begin{equation}
\chi^2 = \sum_l (C_l^{\rm approx} - C_l^{\rm num})^2 
(\sigma_l^{\rm MAP})^{-2}
\label{error_stat}
\end{equation}
where $(\sigma_l^{\rm MAP})^2$ is the variance of $C_l$ in our
model MAP data. For MAP, the measured power spectrum will consist of
approximately 800 independent multipole moments, so an approximate
power spectrum which differs from the actual model power spectrum by
the MAP error at each multipole will have a value for the statistic
$\chi^2$ of around 800.  Figure \ref{error_plot} shows the
error distribution for our subset of models. The statistic $\chi^2$
for each model is plotted against the ``distance'' in physical
parameter space of the model from the fiducial model, where the
distance measure is the Fisher matrix:
\begin{equation}
d = \left[({\bf P}-{\bf P}_0)F({\bf P}-{\bf P}_0)\right]^{1/2},
\label{param_distance}
\end{equation}
where ${\bf P}$ is a vector of physical parameters corresponding to
the model considered, ${\bf P}_0$ is the vector of fiducial-model
parameters, and 
\begin{equation}
F_{ij} = \sum_l {1\over\left(\sigma_l^{\rm MAP}\right)^2}
{\partial C_l\over \partial P_i}{\partial C_l\over \partial P_j}
\label{fisher_matrix}
\end{equation}
is the Fisher matrix. This is roughly equivalent
to measuring the distance along each axis in the physical parameter
space in units of the 1-$\sigma$ error interval in that parameter direction.
The solid-circle points are models with
$2 < N_\nu < 4$, for which the worst-approximated
models disappear. The difference has two possible origins: either
linear extrapolation is not valid over a large enough range in the
variables $\cal R$ and $\cal M$ to handle these models, or CMBFAST
does not provide accurate computations in this range. The second
possibility seems more likely, since the code contains an explicit
trap preventing computations for these unusual numbers of neutrino
species. 

For essentially all other models, the error averaged over all
multipoles is small compared to the MAP statistical errors, and the
error distribution is roughly independent of the distance in parameter
space, showing that our numerical approximations are valid.
Furthermore, most of the total error tends to come from a small number
of multipoles, indicating that systematic errors in CMBFAST dominate
the total difference between our approximations and the direct
numerical calculations. Given this fact, we forego a more thorough
characterization of the errors in our approximate spectrum
computations until a more accurate code is available, but 
the approximations are likely accurate enough for reliable parameter
analysis. It is unclear at this point whether the errors in our
approximations or the systematic errors in CMBFAST will have a greater
impact on the inferred error region.
(Several other public CMB Boltzmann codes are in fact
available, but none are as general as CMBFAST, allowing for arbitrary
curvature, vacuum energy, and neutrino species, all of which are
required for the analysis presented here.)

\section{Discussion and Conclusions}
\label{sec:discussion}

Error regions similar to Fig.~\ref{errors_cosmological_full} have been
previously constructed for model microwave background experiments
\cite{bon99,lew02}. The remarkable point about our calculation is that the
entire error region, constructed from power spectra for $3\times 10^4$
points in
parameter space, has been computed in a few seconds of time on a desktop
computer. In fact, since our power spectra can be computed with a few
arithmetic operations per multipole moment, the calculation time might
not even be dominated by computing model power spectra, but rather by
computing the likelihood function or by converting between the
cosmological and physical parameters.

For actual non-diagonal covariance matrices describing real
experiments, the computation of the likelihood will dominate the total
computation time.  Gupta and Heavens have recently formulated a method
for computing an approximate likelihood with great efficiency
\cite{gup01}, based on finding uncorrelated linear combinations of the
power spectrum estimates \cite{hea00}. Each parameter corresponds to a
unique combination, so the likelihood is calculated with a handful of
operations. Moreover, the method is optimized so the parameter
estimation can be done with virtually no loss of accuracy. Using
this technique, likelihood estimates for realistic covariance
matrices will be roughly as efficient as our power spectrum
estimates. 

The conversion from the physical parameters to cosmological ones
requires evaluation of several numerical integrals. The integrands
will be smooth functions, and this step can likely be made nearly as
efficient as the power spectrum evaluation, although we have not
implemented an optimized routine for parameter conversion.  We make
the following rough timing estimate: each multipole of the power
spectrum can be approximated by a handful of floating point
operations, and an approximate likelihood and the parameter conversion
both will plausibly require similar computational work.  So we
anticipate on the order of 10 floating point operations per multipole
for each of 1000 multipoles.  On a 1 gigaflop machine,
this results in evaluation of around $10^5$ models per second, so a
Markov Chain with tens of thousands of models can be computed in under
a second.  Of course, such a computation is trivially
parallelizable, and can easily be sped up by a factor of tens on a
medium-size parallel machine.  In contrast, other state-of-the-art
parameter estimation techniques \cite{kno01,kap02,spergel}, dominated by the
power spectrum calculation, compute roughly one model per second,
requiring several hours for a $10^4$ point computation.

This great increase in computational speed is highly useful.  All
upcoming microwave background measurements and resulting parameter
estimates will be dominated by systematic errors, both from
measurement errors when observing the sky and processing errors in the
data analysis pipeline.  The only way to discern the effect of these
errors is through modelling them, requiring a determination of
parameters numerous times. The microwave background has the potential
to constrain parameters at the few percent level.  But a variation of
this size in any given parameter will result in changes in $C_l$'s of
a few percent, which is significantly smaller than the errors with
which each $C_l$ will be measured.  This means that small systematic
effects distributed over many Cl's can bias derived parameter values
by amounts larger than the formal statistical errors on the
parameter. Exhaustive simulation of a wide range of potential
systematics will be required before the accuracy of highly precise
parameter determinations can be believed, and our techniques make such
investigations far faster and easier.

Along with increased speed, our approximation methods also promise
high accuracy. A careful error evaluation is hindered by systematic
errors in CMBFAST; our numerical approximations give better accuracy
than the predominant numerical code for parameter ranges generally
larger than the region which will be allowed by the MAP data. The
physical set of variables presented here clarify the degeneracy
structure of the power spectrum, clearly displaying degenerate and
near-degenerate directions in the parameter space.  For example, with
the physical parameters presented here, it is clear that any tensor
mode contribution, which affects only low-$l$ multipoles, will have
negligible impact on the error contours in
Fig.~\ref{errors_physical_full}.

Bond and Efstathiou used a principle component analysis to claim that
uncertainties in the tensor mode contributions would be the dominant
source of error for all of the cosmological parmeters, and further
claimed that as a result, the Fisher matrix approximation will
overestimate the errors in other apparently well-determined parameters
such as our ${\cal B}$ \cite{bon99}. These results are clearly valid
only for data which is significantly less constraining than the MAP
data will be. Essentially, their approximate degeneracy trades off
variations in $\cal R$ and $\cal B$, which change the heights of the
first few peaks, with variations in $n_s$, which also impact the first
few peaks, while holding $\cal A$ fixed.  From our analysis, it is
clear that this degeneracy only holds approximately, since $\cal R$
and $\cal B$ have a significant impact only on the first three peaks,
while $n_s$ affects all of the peak heights.  Indeed, a careful
examination of Fig.~3 in \cite{efs02} shows that the power spectra are
only approximately degenerate, with peak height differences which are
significant for MAP. The addition of tensor modes allows a rough match
between the two power spectra at low $l$ values, but for high $l$
values the discrepancy in peak heights is already evident at the third
peak and will become larger for higher peaks, since the values of
$n_s$ in the two models are so different.

Gravitational lensing makes a negligible difference in the linearity
of the parameter dependences. We have not included polarization in
this analysis, but the three other polarized CMB power spectra can be
handled in the same way as the temperature case using standard
techniques \cite{kam97,zal97c}.

We have shown that a proper choice of physical parameters enables the
microwave background power spectrum to be simply approximated with
high accuracy over a significant region of parameter space. This
region will be large enough for analyses of data from the MAP
satellite, although larger regions could be stitched together using
multiple reference models.  Our approximation scheme requires first
computing numerical derivatives of the power spectrum multipoles with
respect to the various parameters, but this must be done only once and
then computing further spectra is extremely fast (in the neighborhood
of $10^5$ models per second or more on common computers).  The error
determination techniques in this paper require the numerical
evaluation of only tens of power spectra rather than thousands or
millions, so speed of the power spectrum code will not be of paramount
concern. On the other hand, we emphasize that for {\it any}
high-precision parameter estimate, even small systematic errors in
computing $C_l$'s can lead to biased parameter estimates comparable to
the size of the statistical errors. These two considerations argue for
a significant revision of CMBFAST (CMBSLOW, perhaps?)  or construction
of other independent codes which focus on overall accuracy and
stability of derivatives with respect to parameters rather than on
computational speed. Such a code, combined with the estimation
techniques in this paper and efficient likelihood evaluation methods
\cite{gup01} will provide a highly efficient and reliable way to
constrain the space of fundamental cosmological parameters.

\begin{acknowledgments}
We thank David Spergel, who has independently implemented some of the
ideas in this paper, for helpful discussions, and Lloyd Knox, Manoj
Kaplinghat, and Alan Heavens for useful comments. We also thank Uros
Seljak and Matias Zaldarriaga for making public their CMBFAST code,
which has proven invaluable for this and so much other cosmological
analysis. A.K. has been supported in part by NASA through grant
NAG5-10110, and is a Cotrell Scholar of the Research Corporation.
\end{acknowledgments}

\appendix

\section{A Model Pipeline for Parameters Analysis}
\label{app:pipeline}

In the interests of completeness, in this Appendix we outline an
analysis pipeline for cosmological parameters and their errors.  Our
basic data set will be a high-resolution microwave background map,
from which has been extracted an angular power spectrum (i.e., the
$C_l$'s) and a covariance matrix for the $C_l$'s. Given this data set,
plus potentially relevant data from other sources, we need to find the
best-fit cosmological model and the error contours in the parameter
space.

In the case of an ideal, full-sky map with only statistical errors and
uniform sky coverage, the covariance matrix reduces to a multiple of
the identity matrix, but in general, for most experiments the coviance
matrix may contain significant off-diagonal terms. Even for the MAP
satellite, which will come close to a diagonal covariance matrix, the
need to mask out the galactic plane induces some correlations between
nearby multipoles. The first step in the analysis is to construct a
likelihood calculator: what is the likelihood that the measured data
represents some particular theoretical power spectrum? This is in
general a computationally hard problem, since it involves inverting an
$N\times N$ matrix, where $N$ is the number of multipoles in the power
spectrum.  However, the problem becomes easier if the covariance
matrix is close to diagonal. Heavens and Gupta have recently
formulated a method for computing an approximate likelihood with great
efficiency \cite{gup01}, based on an eigenvector decomposition of the
covariance matrix \cite{hea00}. They have demonstrated a likelihood
calculator which takes an input theoretical power spectrum and outputs
a likelihood based on simulated measurements and covariances in much
less than a second of computation time. Further efficiency
improvements are desirable, as this is likely to be the piece of the
analysis pipeline which dominates the total time.

The likelihood calculator can also incorporate prior probabilities
obtained from other data sources, or from physical considerations like
$\Lambda > 0$. Since none of the following steps depend on the form of
the likelihood function, the inclusion of other kinds of data can be
handled in a completely modular way.

The second element needed is a pair of routines to convert between the
physical parameter space (${\cal A}$, ${\cal B}$, ${\cal M}$, ${\cal
V}$) and the cosmological parameter space ($\Omega_{\rm mat}$,
$\Omega_b$, $\Omega_\Lambda$, $h$). To go from the cosmological
parameters to the physical parameters simply involves evaluating the
integral needed in Eq.~\ref{parameterA}. The reverse direction can be
done straightforwardly by rewriting ${\cal A}$ in terms of ${\cal B}$,
${\cal M}$, ${\cal V}$, ${\cal R}$, and $h$ and then numerically
finding the value of $h$ which gives the needed value of ${\cal A}$.
The total variation of $\cal A$ with $h$ is a smooth function, and the
root-finding step can be done with a minimal number of integral
evaluations.  This procedure does not guarantee a solution, however,
in extreme regions of parameter space. A more general (but slower)
method is to search the cosmological parameter space for the model
which best fits the given set of physical parameters, and then project
this model back into the physical parameter space to check the quality
of the fit.

Once a likelihood calculator and parameter conversion routines are in
hand, we advocate the following procedure for parameter
determination. First, choose a fiducial model which gives a good fit
to the measured power spectrum. This can largely be done by eye, using
the parameter dependences displayed in the Figures in
Sec.~\ref{sec:power_spectra}.  Note that the fiducial model does not
need to be the formal best-fit model, but only needs to be close
enough to the best-fit model so that models throughout the error
region will be accurately approximated by the extrapolations and
approximations in Sec.~\ref{sec:power_spectra}.

Next, check the validity of the fiducial model. Is the fit of the
model with the data adequate? It is certainly possible that the simple
space of cosmological models considered here may not contain the
actual universe. For example, some admixture of isocurvature initial
conditions might be important, or the initial power spectrum might not
be well described by a power law. If the best-fit model is not a good
fit to the data, it makes no sense to proceed with any further
parameter analysis! The model space must be enlarged or modified to
include models which provide a good fit to the data.

If the fiducial model is adequate, then compute numerical partial
derivatives of each $C_l$ with respect to ${\cal B}$, ${\cal M}$, and
${\cal R}$. Also, write numerical routines to compute the power
spectrum approximately for $l<50$. Even though CMBFAST or similar
codes run very efficiently when computing only $2<l<50$, this time
will still greatly dominate the total computation time for parameter
error determination, so an efficient analytic estimate is
essential. Note that an alternate possibility is to simply ignore the
lowest multipoles, which will be a small portion of the data for a
high-resolution map, at the expense of losing any leverage on the
parameters ${\cal V}$, ${\cal Z}$, and the tensor power spectrum.
Still, the other parameters will be determined much more precisely, so
it may be possible to fit the rest of the parameters using $l>50$ data
only, and then constrain just the ${\cal V}$-${\cal Z}$-tensor
parameter space from the $l<50$ data independently. Now it is possible
to estimate the $C_l$'s highly efficiently for any model within a
sizeable parameter space region around the fiducial model.

Now perform a Markov-chain Monte Carlo \cite{chr01} in the physical
parameter space.  Each point in parameter space requires evaluation of
one approximate power spectrum, one likelihood, and one conversion to
cosmological parameters.  The Monte Carlo will converge efficiently
since the physical parameters are nearly orthogonal in their effects
on the power spectrum.  Sophisticated convergence diagnostics are
available for determining when the Markov chain of models has
sufficiently sampled the error region \cite{gil96}.

Finally, once the chain of models in parameter space has been
computed, extract the best-fit model and error contours from the
distribution of models.  By displaying two-dimensional likelihood
plots for all parameter combinations, a reliable picture of the shape
of the likelihood region in the whole multidimensional parameter space
can be visualized.

Upcoming data will be good enough that the statistical errors, given
by these likelihood contours, will be smaller than the systematic
errors, at least in some parameter space directions.  Sources of
systematic error include instrumental effects, modelling of foreground
emission, approximation of the covariance matrix, inaccuracies in the
theoretical models, and inaccuracies in the approximations in this
paper. Before we can confidently claim to have determined the
cosmological parameters and their errors, these systematics must be
modelled and investigated. Using the techniques outlined here, where a
complete Monte Carlo can be accomplished in seconds, investigation of
systematics is easily practical, instead of a computational challenge.

\section{Some Numerical Considerations with the CMBFAST Code}
\label{app:numerics}

Seljak and Zaldarriaga's CMBFAST code \cite{sel96,zal00} has become the
standard tool for computing the microwave background temperature and
polarization power spectra for the inflation-type cosmological models
discussed in this paper. Hundreds of researchers have employed it for
tasks ranging from predictions of highly speculative cosmological
models to analysis of every microwave background measurement of the
past few years. Its comprehensive treatment of a variety of physical
effects and its public availability have made it the gold standard for
microwave background analysis.

The original aim of the CMBFAST code was fast evaluation of the power
spectra. Previous codes had simply evolved the complete Boltzmann
hierarchy of equations up to the maximum desired value of $l$. This
resulted in a coupled set of around $3l$ linear differential
equations, and the $l$-values of interest might be 1000 or
greater. The task of evolving thousands of coupled equations with
oscillatory solutions led to codes which required many hours to run on
fast computers. The CMBFAST algorithm evolves the source terms first,
which involve angular moments only up to $l=4$; the same Boltzmann
hierarchy can be used, but with a cutoff of $l\simeq 10$.  The rest of
the $C_l$'s can then be obtained by integrating this source term
against various Bessel functions. The Bessel functions can be
precomputed or approximated \cite{kos98} independently for each
$l$. Great time savings result since (1) it is not necessary to
compute the power spectrum for every $l$, but rather only, e.g., every
50th $l$ with the rest obtained from interpolation; (2) the time steps
for the Boltzmann integration are not controlled by the need to
resolve all of the oscillations in the Bessel functions, and thus the
Boltzmann integrator can be significantly more efficient. It also may
be possible to employ approximations to the integrals of the source
times the Bessel functions with an additional gain in speed, although
CMBFAST does not implement this. The code offers a speed improvement
of a factor of 50 over Boltzmann hierarchy codes for flat universes,
although the advantage is smaller for models which are not flat or
have a cosmological constant.

CMBFAST is nominally accurate at around the one percent level,
although no systematic error analysis has ever been published.  While
the numerical error in any given multipole $C_l$ is relatively small,
the derivatives of $C_l$ with respect to the various cosmological
parameters are significantly less accurate.  At the time the code was
written, no microwave background experiments were of sufficient
precision to warrant worrying about inaccuracies at this level, and
the great utility of the code has been its combination of speed and
accuracy. But we have now entered the era of high-precision
measurements which will place significant constraints on many
parameters simultaneously, and as this paper has demonstrated,
efficiently prosecuting this analysis requires a code of high
precision and stable behavior with respect to parameter variations.
We have found that the unmodified version of CMBFAST ``out of the
box'' is not sufficiently accurate for the calculations presented
here, but these shortcomings are partially compensated by the
following procedures.

First, CMBFAST uses slightly different algorithms for flat and
non-flat background spacetime geometries. In the flat case, the
necessary spherical Bessel functions are precomputed and called from a
file; in the nonflat case, the hyperspherical Bessel functions are
computed via an integral representation as they are needed. As a
result, the line-of-sight integrals over the product of source term
and Bessel function have different partitions in the two cases, and
the numerical values differ at the percent level. Also, the code has a
flag which forces evaluation of the flat case whenever $|\Omega
-1|<0.001$.  If the fiducial model is flat, then as parameters are
varied which change the geometry, many individual multipoles $C_l$
exhibit discontinuities at the point in parameter space where the code
switches from the flat to nonflat evaluation scheme. These
discontinuities can result in significant errors when evaluating $C_l$
derivatives with respect to parameters. Also, even if the derivative
is correctly evaluated, the offset between the two cases can result in
a slightly biased error estimate.

A careful fix of this problem would involve insuring that the integral
partitions used in the two cases are determined consistently.
Alternately, the flat-space code could be rewritten to numerically
integrate the necessary Bessel functions as in the non-flat case,
since the two methods do not have a significant difference in
computational speed.  A simpler practical solution which we have
employed in this paper is to force the code always to use the non-flat
integration routine.  This can be accomplished by narrowing the
tolerance at which the code uses the non-flat integration (say to
$|\Omega -1|< 10^{-5}$), combined with always using a non-flat
fiducial model (shifting $\Omega_{\rm mat}$ by $10^{-5}$ will never
result in a statistically significant difference when fitting a given
data set, due to cosmic variance).

Second, in varying the ${\cal B}$ parameter, adjacent acoustic peaks
move in the opposite direction. Between the peaks, therefore, is a
pivot point $l$ for which $C_l$ remains fixed as ${\cal B}$ is varied,
and the nearby $C_l$'s will vary only slightly. As CMBFAST is written,
it computes $C_l$ at $l$ values separated by 50; the complete spectrum
is then obtained by splining between the calculated values. Splines
are rather stiff, and small changes in regions where the power
spectrum has significant curvature (i.e. around the acoustic peaks and
troughs) can noticeably change the overall fit to the power spectrum
in the regions between. In particular, the position of the ``pivot''
$l$-value in the spectrum sometimes shifts a bit, which gives a
spurious irregular dependence on ${\cal B}$ of the few $C_l$'s in this
region.

\begin{figure}
\includegraphics[width=3.4in]{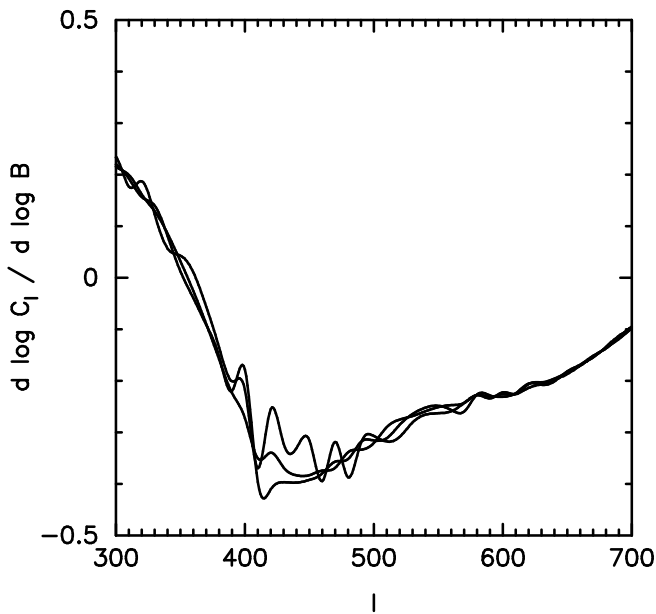}
\caption{Derivatives of $C_l$ with respect to ${\cal B}$, calculated
by CMBFAST with $\Delta l = 10$,
from 2\%, 5\%, and 10\% finite differences in ${\cal B}$. The smoothest
curve is for the 10\% variation.}
\label{cmbfast_problem}
\end{figure}

This can be addressed by a simple modification to CMBFAST forcing the
explicit evaluation of $C_l$ for more $l$ values. We have used a
$\delta l$ of 10, which has eliminated the spurious behavior. Of
course, this results in a significant increase in the computational
time needed for an individual model.  A more sophisticated approach
would be to use a comparable number of $l$ values as the original
code, but distribute them preferentially in the $l$-ranges
corresponding to the peaks and troughs where the curvature of the
power spectrum is largest. This is easily implemented using the
variables defined in this paper, since the peak positions are
completely determined by ${\cal A}$ alone. Using a finer grid in $l$
also reproduces the peak positions with higher accuracy.

Third, as discussed in the body of the paper, the power spectra must
be normalized according to a physical criterion which does not depend
on non-linear behavior at small $l$ values.  This is accomplished by
simply compiling CMBFAST with the UNNORM option. In this case, no
normalization to COBE is done; instead, the amplitude of the potential
$\Psi$ is taken to be unity at the scale of the horizon at the time of
last scattering.  This normalization is fixed by the primordial
amplitude of scalar perturbations, often denoted as $A_S$.  We define
a physical variable $S$ corresponding to the microwave background
amplitude by also including the effect of differing amounts of
reionization so that the power spectrum at large $l$ remains fixed.

Some residual numerical problems remain with CMBFAST.  We have found,
for example, that in the region around $l=350$ (between the first and
second acoustic peaks for our fiducial model), the power spectrum
exhibits spurious oscillatory behavior with an amplitude of a few
percent, if computed directly for every $l$ instead of every 10th or
50th $l$ (see Fig.~\ref{cmbfast_problem}). 
This likely arises from either the $k$ or $\eta$ integral
over an oscillatory integrand not being evaluated on a fine enough
grid. As a result, directly calculated $C_l$ values in this
region of the parameter space effectively have non-negligible
random errors. This problem is normally hidden by the spline smoothing with
every 50th $l$ computed. When the number of $C_l$ evaluations is
increased to every 10th $l$, the power spectrum shows a distinct
glitch in this region, and the parameter dependence is completely
unreliable here. The spline fits are forced into
more oscillations to include the computed points.
More accurate numerical integrations will probably
solve this problem, at the expense of a significant speed reduction.

\bibliography{params}

\end{document}